\documentclass[11pt]{article}

\pdfoutput=1
\usepackage{latexsym}
\usepackage{epsfig}
\usepackage[mathscr]{eucal}
\usepackage{amsfonts}
\usepackage{amscd}
\usepackage{cite}
\usepackage{array}
\usepackage{amssymb}
\usepackage{colordvi}
\usepackage[centertags]{amsmath}
\usepackage{enumerate}
\usepackage{graphicx}
\usepackage{booktabs}
\usepackage{theorem}
\usepackage[footnotesize]{caption}
\usepackage{soul}
\usepackage{url}
\usepackage{slashed}
\usepackage{color}
\usepackage{bbm}
\usepackage[utf8]{inputenc}

\def\be{\begin{equation}}
\def\ee{\end{equation}}
\def\ba{\begin{eqnarray}}
\def\ea{\end{eqnarray}}

\def\ltap{\;\centeron{\raise.35ex\hbox{$<$}}{\lower.65ex\hbox{$\sim$}}\;}
\def\gtap{\;\centeron{\raise.35ex\hbox{$>$}}{\lower.65ex\hbox{$\sim$}}\;}

\def\half{\tfrac{1}{2}}
\newcommand{\bea}{\begin{eqnarray}}
\newcommand{\eea}{\end{eqnarray}}

\makeatletter
\def\section{\@startsection {section}{1}{\z@}{-3.5ex plus -1ex minus -.2ex}{2.3ex plus .2ex}{\large\bf}}
\def\subsection{\@startsection{subsection}{2}{\z@}{-3.25ex plus -1ex
minus -.2ex}{1.5ex plus .2ex}{\normalsize\bf}}
\makeatother
\newcommand{\captionfonts}{\small}
\makeatletter  
\long\def\@makecaption#1#2{%
  \vskip\abovecaptionskip
  \sbox\@tempboxa{{\captionfonts #1: #2}}%
  \ifdim \wd\@tempboxa >\hsize
    {\captionfonts #1: #2\par}
  \else
    \hbox to\hsize{\hfil\box\@tempboxa\hfil}%
  \fi
  \vskip\belowcaptionskip}
\makeatother   

\catcode`@=11
\def\marginnote#1{}
\newcount\hour
\newcount\minute
\newtoks\amorpm
\hour=\time\divide\hour
by60
\minute=\time{\multiply\hour by60 \global\advance\minute
by-\hour}
\edef\standardtime{{\ifnum\hour<12 \global\amorpm={am}
\else\global\amorpm={pm}\advance\hour by-12 \fi
 \ifnum\hour=0
\hour=12 \fi
 \number\hour:\ifnum\minute<10
0\fi\number\minute\the\amorpm}}
\edef\militarytime{\number\hour:\ifnum\minute<10
0\fi\number\minute}
\def\draftlabel#1{{\@bsphack\if@filesw
{\let\thepage\relax
 \xdef\@gtempa{\write\@auxout{\string
\newlabel{#1}{{\@currentlabel}{\thepage}}}}}\@gtempa
 \if@nobreak
\ifvmode\nobreak\fi\fi\fi\@esphack}
\gdef\@eqnlabel{#1}}
\def\@eqnlabel{}
\def\@vacuum{}
\def\draftmarginnote#1{\marginpar{\raggedright\scriptsize\tt#1}}
\def\draft{\oddsidemargin
0.0truein
 \def\@oddfoot{\sl preliminary draft \hfil
\rm\thepage\hfil\sl\today\quad\militarytime}
 \let\@evenfoot\@oddfoot
\overfullrule 3pt
 \let\label=\draftlabel
\let\marginnote=\draftmarginnote
\def\@eqnnum{(\theequation)\rlap{\kern\marginparsep\tt\@eqnlabel}
\global\let\@eqnlabel\@vacuum}
}
\catcode`@=12

\newcommand{\beq}{\begin{eqnarray}}
\newcommand{\eeq}{\end{eqnarray}}

\begin{document}

\thispagestyle{empty}

\begin{center}

\begin{center}

\vspace{1.7cm}

{\LARGE\bf Pseudoscalar decays to gauge bosons at the LHC and at a future 100 TeV collider}
\end{center}

\vspace{1.4cm}

\renewcommand{\thefootnote}{\fnsymbol{footnote}}
{\bf Abdesslam Arhrib$^{\,1\,}$}\footnote{E-mail: \texttt{aarhrib@gmail.com}},
{\bf Rachid Benbrik$^{\,2}$}\footnote{E-mail: \texttt{r.benbrik@uca.ac.ma}},
{\bf Jaouad El Falaki$^{\,1,\, 3}\,$}\footnote{E-mail: \texttt{jaouad.elfalaki@gmail.com}},
{\bf Marco Sampaio$^{\,4,\, 5 \,}$}\footnote{E-mail: \texttt{msampaio@ua.pt}} and
{\bf Rui Santos$^{\,6,\, 7}$}\footnote{E-mail: \texttt{rasantos@fc.ul.pt}}
\\

\vspace{1.cm}

${}^1\!\!$
{\em D\'epartement de Math\'ematiques, Facult\'e des Sciences et Techniques,} \\
{\em Universit\'e Abdelmalek Essaadi, B. 416, Tangier, Morocco}
\\
${}^2\!\!$
{\em LPHEA, MSISM Team D\'epartment of Physics,} \\
{\em Facult\'e Polydisciplinaire de Safi, Sidi Bouzid, BP 4162,  Safi, Morocco}
\\
${}^3\!\!$
{\em Center  for  Fundamental  Physics,  Zewail  City  of  Science  and  Technology,} \\
{\em  Sheikh  Zayed,12588 Giza, Egypt,}
\\

${}^4\!\!$
{\em Departamento de F\'{\i}sica da Universidade de Aveiro,} \\
{\em Campus de Santiago, 3810-183 Aveiro, Portugal}
\\
${}^5\!\!$
{\em CIDMA - Center for Research \& Development in Mathematics and Applications,} \\
{\em Campus de Santiago, 3810-183 Aveiro, Portugal}
\\
${}^6\!\!$
{\em {ISEL - 
 Instituto Superior de Engenharia de Lisboa,\\
 Instituto Polit\'ecnico de Lisboa 
 1959-007 Lisboa, Portugal}\\
${}^7\!\!$
{\em Centro de F\'{\i}sica Te\'{o}rica e Computacional,
    Faculdade de Ci\^{e}ncias,
    Universidade de Lisboa,} \\
{\em Campo Grande, Edif\'{\i}cio C8 1749-016 Lisboa, Portugal}
}
\end{center}

\vspace{1.8cm}
\centerline{\bf Abstract}
\vspace{2 mm}
\begin{quote}
\small 
We discuss the search for a CP-odd scalar decaying into gauge bosons
in the framework of a CP-conserving two-Higgs doublet model (2HDM) and 
of a 2HDM extended with a vector-like quark (VLQ) at the Large Hadron Collider
and at a future 100 TeV collider. The rate of decay of a pseudoscalar 
to $Z$-bosons could be important to ascertain the CP-nature of the scalars
in the model. In the 2HDM $A \to ZZ$ will be extremely
hard to detect even at a future 100 TeV $pp$ collider while in the 2HDM+VLQ 
this decay can be probed even during the present LHC run. We further discuss all decays
of the pseudoscalar into gauge bosons at the LHC and at a future 100 TeV collider
in the alignment limit where the lightest scalar is the 125 GeV Higgs with SM-like couplings
to the fermions and gauge bosons.

\end{quote}

\newpage
\setcounter{page}{1}
\setcounter{footnote}{0}

\section{\label{sec:intro} Introduction}

After the discovery of the Higgs boson by the ATLAS~\cite{Aad:2012tfa} 
and CMS~\cite{Chatrchyan:2012xdj} collaborations 
at the Large Hadron Collider (LHC) the high energy physics community 
focused on the search for signs of extended
scalar sectors~\cite{deFlorian:2016spz}. Such extended sectors, with extra 
Higgs singlets, doublets or triplets, are a common
feature of several Beyond the Standard Model (BSM) models.
Finding a new scalar would be a clear signal of BSM physics with extended Higgs sectors.
Two-Higgs doublet models (2HDM)~\cite{Lee:1973iz}, both in their
CP-conserving and CP-violating versions, have been used as benchmark models to
search for new scalars at the LHC. The 2HDM has four extra degrees of freedom 
with two extra neutral scalars and two charged scalars. 
In the CP-conserving version of 
the model the two neutral states are $h$ and $H$ (CP-even)
and $A$ (CP-odd) while in the CP-violating version the three neutral states
are a mixture of CP-even and CP-odd states and are referred to as $h_1$,
$h_2$ and $h_3$. In this work we focus on the production of the 
pseudoscalar via gluon fusion plus $b \bar b$ initiated
process with the subsequent decay to gauge bosons, with focus on 
$A \to \gamma \gamma$ and $A \to ZZ$. Although
$A \to \gamma \gamma$ is a loop induced process, it is 
nevertheless competitive with other final states like $\tau^+ \tau^-$ 
in large portions of the parameter space of the model. Hence, a pseudoscalar 
could first be detected in the two photon final
state. However, even if $A$ is discovered in some other final states, the 
remaining possible decays have either to be confirmed
or excluded. This study will therefore give us further information on the
model.

Previous works have discussed the pseudoscalar 
decays into gauge bosons  in a variety of models such as the
2HDM~\cite{Arhrib:2006rx, Diaz-Cruz:2014aga, Chowdhury:2017aav}, the 2HDM with a sequential fourth
generation of quarks~\cite{Bernreuther:2010uw} and in Supersymetric 
Models~\cite{Gunion:1991cw}.
 In this work, our main focus either 
diverges or completes the previous studies. In terms of completing and/or updating the studies we discuss
how both $A \to \gamma \gamma$ and $A \to ZZ$ are affected by the latest experimental searches at the LHC
together with the most relevant and up-to-date experimental and theoretical constraints.
We then move to discuss what is expected by the end of the LHC and also at a 
future 100 TeV pp collider. We include for the first time the study of the processes in an extension of the 2HDM 
with the addition of vector like quarks (VLQ) \cite{Aguilar-Saavedra:2013qpa, Cacciapaglia:2015ixa, Arhrib:2016rlj}.
The extra quark loops present both in the production and in the decay may lead to a significant enhancement of the 
rates relative to the 2HDM. Therefore, the rate $pp \to A \to ZZ$ can differ
by several orders of magnitude in the 2HDM and in the 2HDM+VLQ which shows that loop processes
can vary by orders of magnitude in two simple extensions of the SM. This extension of the 2HDM is used to show that the range of possible
variation of the number of events produced at loop-level can be very similar to the ones
produced at tree-level in a model with the same potential but with extra fermion content. 

Most importantly, a question that was not asked in previous works is: to what extent can we say
if we are indeed seeing a pseudoscalar?  We want to understand what to expect if a scalar is found decaying into a $ZZ$ final state. 
Therefore, not only we discuss the event rates of a $pp \to A \to ZZ$ in two different models, but also discuss the different possibilities
of distinguishing a scalar with definite CP from a CP-violation one in the context of 2HDMs.

ATLAS and CMS have shown that if the 125 GeV Higgs has a definite CP, then it does not have $CP=-1$.
The search for CP-violation in the  $HZZ$ vertex was and is still being performed at the LHC~\cite{Aad:2013xqa, Sirunyan:2017tqd}
using the method described in~\cite{Choi:2002jk, Buszello:2002uu}. However, it is very important to stress that
in models such as CP-violating versions of the 2HDM, as in all CP-violating extensions with singlets and doublets,
the $HZZ$ has its origin in the Lagrangian term $(D_\mu \phi)^\dagger \, D^\mu \phi$, where $\phi$ is an $SU(2)$ doublet, and therefore
it has the simple SM Lorentz structure proportional to $g_{\mu \nu}$ at tree-level. Therefore the measurements of the effective operators
in~\cite{Aad:2013xqa, Sirunyan:2017tqd} can at most to be used to constraint these models at loop-level. 

One way to look for signs of CP-violation at the LHC is in the decays to $ZZ$~\cite{Fontes:2015xva}. Several classes of processes may hint a signal of CP-violation. One
such example is the combined observation of the three decays $h_2 \to h_1 Z$, $h_2 \to Z Z$ and $h_1 \to Z Z$, where $h_1$ is the 125 GeV Higgs. Except for the already measured $h_1 \to Z Z$, the other two processes
that occur at tree-level in a CP-violating model, can be mistaken by the loop processes 
$A \to h Z$ and $A \to ZZ$ from the corresponding CP-conserving model. In the alignment limit, where all $h_1/h$ couplings to the
SM particles mimic the SM ones, $A \to h Z$ is zero at tree-level. Also, $A \to ZZ$ is a loop
induced process and therefore very small. Hence, if a new scalar is found, its CP-numbers
could be hard if not at all impossible to determine with this method if the rates are too small. 

Another way to look for a similar effect in a single process is to look for anomalous (CP-violating) coupling in the 
triple gauge bosons vertices~\cite{Gaemers:1978hg, Hagiwara:1986vm, Gounaris:1999kf, Gounaris:2000dn}.
The structure of the off-shell $Z^*ZZ$ vertex, has terms that are only non-zero if CP-violation is present in the model. This 
was calculated for the particular case of a complex version of the 2HDM (the C2HDM)~\cite{Grzadkowski:2016lpv, Belusca-Maito:2017iob} 
(see~\cite{Fontes:2017zfn} for a recent review on the C2HDM)
and also for an extension of the 2HDM with an extra singlet where CP-violation
only exists in the dark scalar sector~\cite{Azevedo:2018fmj} and therefore looking for the three decays that compose
the one loop contribution to $Z^*ZZ$ is not an option. In this model, the only loop diagram that contributes 
to $Z^*ZZ$ involves all vertices $h_i ZZ, \, i=1,2,3$.
The present LHC results probe the CP-violating term $f_4^Z$ to order  $\approx 10^{-3}$~\cite{Aad:2012awa, CMS:2014xja, Khachatryan:2015pba}, 
whereas the typical magnitude of the $f_4^Z$ term (both real and imaginary parts) is $\approx 10^{-5}$.  Recently, a study for the LHC~\cite{Rahaman:2018ujg}
has shown that for the LHC@14TeV an increase in luminosity from 36$fb^{-1}$ to 1000$fb^{-1}$ implies an improvement on the measurement by only a factor of 2, even
with the inclusion of asymmetries which were shown not to have a significant impact in the limit. Therefore, the typical magnitudes will probably be out of reach
even for the high luminosity stage, although no study is still available. There are also no studies for a future 100 TeV $pp$ collider.

The previous discussion shows that probing the CP-nature of a new scalar is going to be extremely hard. In fact, if a new scalar is found in the $ZZ$ channel, especially with low rates,
we believe the only possibility to probe its CP-nature is through the Yukawa couplings. In this case the Yukawa couplings have to be large enough to allow this
distinction, which is certainly possible in models like the 2HDM or the C2HDM, where even if $g_{HZZ}$ is small, Yukawa couplings can still be large. The discussion
of CP-violation in the Yukawa couplings will appear in section 7 and in the conclusions.

The paper is organized as follows. In the next section we discuss the 2HDM
setup and list the theoretical and experimental constraints we will be using. In section 
3 we present the 2HDM extended  by an up-type vector-like quark (2HDM+T).
In section 4 we present our results for the 2HDM while in section 5 results
for the 2HDM+T are presented for the LHC Run 2. A discussion on the prospects
for a future 100 TeV collider are examined in section 6. Finally, in section 7 
we address the problem of the contribution of these searches to understand the 
CP nature of a new scalar. Our conclusions are presented in the last section.

\section{The two-Higgs doublet model}
\label{sec:model}

The 2HDM was proposed by T.D.~Lee~\cite{Lee:1973iz} in an attempt to 
explain the matter anti-matter asymmetry of the Universe through 
the addition of an extra source of CP-violation. In this work we discuss
the CP-conserving version of the model that contains
two CP-even states denoted by $h$ (the lightest) and $H$, one
CP-odd state denoted by $A$ and two charged states, $H^\pm$.
As tree-level flavour changing neutral currents (FCNC) are very
constrained by experiments, we impose a $\mathbb{Z}_2$ symmetry
$\Phi_1 \to \Phi_1$ and $\Phi_2 \to - \Phi_2$ on the scalar
fields. The resulting Higgs potential (softly broken by the dimension
two term $m^2_{12}$) can be written as 
\begin{eqnarray} \label{pot}
\mathcal{V} &=& m^2_{11}\Phi_1^\dagger\Phi_1+m^2_{22}\Phi_2^\dagger\Phi_2
  -\left(m^2_{12}\Phi_1^\dagger\Phi_2+{\rm h.c.}\right)
  +\half\lambda_1\left(\Phi_1^\dagger\Phi_1\right)^2
  +\half\lambda_2\left(\Phi_2^\dagger\Phi_2\right)^2 \nonumber \\
&& \qquad +\lambda_3\Phi_1^\dagger\Phi_1\Phi_2^\dagger\Phi_2
  +\lambda_4\Phi_1^\dagger\Phi_2\Phi_2^\dagger\Phi_1
  +\left[\half\lambda_5\left(\Phi_1^\dagger\Phi_2\right)^2+{\rm h.c.}\right]\,.
\end{eqnarray}
Choosing real vacuum expectation values (VEVs), $v_1$ and $v_2$ and demanding
$m_{12}^2$ and $\lambda_5$ to be real as well, the potential is CP-conserving.
One should note that the CP-conserving minimum of any 2HDM is stable at tree-level, that is,
any other stationary point, if it exists, is a saddle point~\cite{Ferreira:2004yd,Barroso:2005sm}.
Still two CP-conserving minima can coexist but the existence of a global minimum
can be easily enforced by a simple condition~\cite{Barroso:2013awa, Ivanov:2008cxa}.   
The free independent parameters are the four masses, $m_h$, $m_H$, $m_A$ and $m_{H^\pm}$,
the soft breaking parameter $m_{12}^2$, the angle $\tan \beta = v_2/v_1$ and the rotation
angle $\alpha$ that diagonalizes the CP-even mass matrix.

When we impose that no tree-level FCNCs are present in 
 the theory by extending the $\mathbb{Z}_2$ symmetry~\cite{Glashow:1976nt, Paschos:1976ay}
to the Yukawa sector, we end up with four independent versions of the model. 
These are: Type I - only $\Phi_2$ couples to all fermions;
Type II - $\Phi_2$ couples to up-type quarks and $\Phi_1$ couples to 
charged leptons and down-type quarks;
Flipped or Type Y - $\Phi_2$ couples to charged leptons and up-type quarks and $\Phi_1$ couples to down-type quarks;
Lepton Specific or Type X -  $\Phi_2$ couples to quarks and $\Phi_1$ couples to charged leptons.


The scan in the 2HDM parameter space was performed fixing $m_h = 125 \, GeV$,  $\sin (\beta - \alpha) = 1$, since a small misalignment has no phenomenological consequences on the CP-odd decay,
and $m_{H^\pm} = m_H = 600 \, GeV$ and varying $m_A$, $\tan \beta$ and $m_{12}^2$ in the allowed
parameter space. This is the exact alignment limit and it is in agreement with the most relevant experimental and
theoretical constraints: 

\begin{itemize}

\item
The potential is bounded from below at tree-level~\cite{Deshpande:1977rw, Ivanov:2006yq};

\item
Perturbative unitarity is enforced~\cite{Branco:2011iw,Kanemura:1993hm, Akeroyd:2000wc} 
to the quartic couplings of the Higgs potential; 
 
\item
The parameter space complies with electroweak precision 
observables~\cite{ALEPH:2010aa} via S and T 
parameters~\cite{Peskin:1991sw, Froggatt:1991qw, Grimus:2008nb,
Haber:2010bw, Baak:2011ze} because $m_{H^\pm} = m_H$ and $\cos (\beta - \alpha) = 0$~\cite{Haber:2010bw};
 
\item
Collider bounds from LEP, Tevatron and from LHC Run 1
are taken into account. Since we work in the alignment limit, $\sin (\beta - \alpha)=1$, automatic agreement with the constraints
on the Higgs couplings to the other SM particles is attained, because all Higgs couplings become SM-like. Regarding the searches,
the tree-level decays to gauge bosons of both $H$ and $A$ are forbidden. The decays to fermions are considered
and of particular relevance are the bounds arising from the search $pp \to A \to \tau^+ \tau^-$~\cite{BruckmanDeRenstrom:2016hbv, Arhrib:2015gra, Sirunyan:2018zut}.
These imply that in the Type II 2HDM, the values of $\tan \beta$ cannot be too large especially for low $m_A$ bounds. Therefore we choose to take $\tan \beta < 10$ in the
entire mass range for Type II. However, one should note that the largest cross sections are the gluon fusion ones, with the maximum value for $\tan \beta =1$. Also the
$BR(A \to \gamma \gamma)$ decreases with $\tan \beta$. Hence, overall the largest rates are obtained for $\tan \beta \approx 1$. We will check directly in our study
how the searches for a pseudoscalar decaying into gauge bosons affects the parameter space.

\item
We consider the most relevant indirect constraints 
on the parameter space in the plane $(m_{H^\pm}, \tan \beta)$.
These are mainly loop processes where cancellations could occur in the loops if other sources
of new physics are considered. The bounds arise mainly from
$B$-physics observables~\cite{WahabElKaffas:2007xd, Aoki:2009ha, Su:2009fz,
Mahmoudi:2009zx, Deschamps:2009rh} and $R_b = \Gamma (Z \to b \bar b)/\Gamma (Z \to hadrons)$
~\cite{Denner:1991ie, Haber:1999zh, Freitas:2012sy}. These constraints result
in a rough bound of $\tan \beta \geq 1$ for all Yukawa types. Regarding the charged
Higgs mass the most relevant bound comes from $b \to s \gamma$ (Type II and Y only)
and is at present $m_{H^\pm} \geq 570$ GeV~\cite{Misiak:2015xwa,Misiak:2017bgg}. 
The same constraints forces $\tan \beta \geq 2$ in Type I and X.
LHC run 1 has contributed
with direct bounds in the $(m_{H^\pm}, \tan \beta)$ plane with
the process $pp \to t ~\bar t (H^+ W^- b \bar b)$~\cite{Chatrchyan:2012vca, Aad:2012tj}.
Finally, LEP provided the only direct bound on the charged Higgs mass of roughly $m_{H^\pm} \geq 90 GeV$
for all Yukawa types with $e^+ e^- \to H^+ H^-$ assuming only 
$BR(H^\pm \to cs) + BR(H^\pm \to \tau \nu) + BR(H^\pm \to AW) = 1$~\cite{Abbiendi:2013hk}.
\end{itemize}

\section{The 2HDM extended by an up-type vector-like quark}
\label{sec:modelvlq}

Vector-like quarks (VLQs) appear naturally in various 
extension of the SM, such as some supersymetric models
\cite{Martin:2009bg}, models with extra-dimensions \cite{Kong:2010qd}, 
little Higgs models \cite{ArkaniHamed:2002qy} and 
composite Higgs models \cite{Kaplan:1983sm}.
VLQs are also well motivated by the fact that
they can solve the Higgs boson mass instability 
resulting from large radiative corrections at high scales. 
In fact, a vector-like top quark partner (T) could play
the same role as the superpartner of the top quark in supersymetric models.
A particular feature of VLQs is that their left and right-handed 
components transform in the same way under the SM gauge group.
Consequently, their mass terms are allowed in the Lagrangian 
without violating gauge invariance transformations.

There have been many studies on the phenomenology of the SM extended with VLQs \cite{Aguilar-Saavedra:2013qpa, Cacciapaglia:2015ixa, Angelescu:2015kga}.  
Moreover, both ATLAS and CMS 
have performed several experimental searches for such new quarks.
Direct searches by the ATLAS and CMS Collaborations have set lower 
limits on the mass of the single vector-like T top partner in the 
range of $550-900$ GeV at $8$ TeV \cite{Aad:2015gdg, Aad:2015kqa, Khachatryan:2015oba} through one of its main decay channels:
 $T\rightarrow ht,Wb, Zt$ since the new top is expected
to couple predominantly to the third generation quarks. 
The above limit was improved in the $13$ TeV run. 
A lower limit on the mass of the T-quark was derived and found to be in the range
 $1170-1295$ GeV by ATLAS and CMS at $13$ TeV \cite{Aaboud:2016lwz, Aaboud:2017qpr, Sirunyan:2017pks, Aaboud:2017zfn}. 
This lower limit can be lowered if the new T-quark has a non-negligible mixing with the first and second
generation quarks \cite{Aad:2012bt}.

In the present study 
we consider an extension of the 2HDM by adding a vector-like top quark (T)
with charge $+\frac{2}{3}$. This extention was already 
studied in detail
in~\cite{Arhrib:2016rlj,Angelescu:2015uiz,Aguilar-Saavedra:2017giu}. 
Similarly to the SM, we introduce 
left and right components of the new top: $T_L^0, T_R^0$.
The 2HDM-VLQ Lagrangian with the new top-quark $T$ is given by:
\begin{eqnarray}
-\mathcal{L}_Y &\supset& y_T\overline{Q_L^0} \widetilde{H}_2T_R^0+
\lambda_T\overline{Q^0_L} \widetilde{H}_1T_R^0+M_T \overline{T_L^0}T_R^0 + \rm{h.c.}\nonumber\\
&=& y_T ( \overline{t_L^0}, \overline{b_L^0})\left(
   \begin{array}{c}
      \frac{\varphi^0_2-iA}{\sqrt{2}} \\
                  -H^- \\
         \end{array}
    \right)T_R^0 +\lambda_T ( \overline{t_L^0}, \overline{b_L^0})\left(
            \begin{array}{c}
       \frac{v+\varphi^0_1-iG^0}{\sqrt{2}} \\
            -G^- \\
      \end{array}
   \right)T_R^0+M_T\overline{T_L^0}T_R^0 + \rm{h.c.},
\end{eqnarray}
where $\tilde H_i \equiv i \tau_2 H^*_i$ and
$Q^0_{L}$ is the left handed third generation quark doublets.
Note that this Lagrangian is valid for all 2HDM-VLQ types because
the couplings to the top are the same in all models. 

After spontaneous symmetry breaking, the top quark mixes with T and the 
mass matrix of the mixing between
$(t_L^0,T_L^0)$ and $(t_R^0,T_R^0)$ is given by (where we can rotate away one off-diagonal element of the mass mixing matrix).
\begin{eqnarray}
{\mathcal M}=\left(
\begin{array}{cc}
    \frac{y_t v}{\sqrt{2}} & \frac{\lambda_T v}{\sqrt{2}} \\
    0 & M_T \\
  \end{array}
        \right),
\label{eq:massmat}
\end{eqnarray}
where $y_t$ and $\lambda_T$ are the Yukawa couplings for the top quark and T.\\
This matrix can be diagonalized by rotating the weak 
eigenstates $(t_L^0,T_L^0)$  into the mass eigenstates $(t_L,T_L)$ using
 a bi-unitarity transformations.
 \begin{eqnarray}
\left(
\begin{array}{c}
 t_{L,R} \\
 T_{L,R}  \\
    \end{array}
\right)={\mathcal U_{L,R}}\left(
\begin{array}{c}
t_{L,R}^0 \\
T_{L,R}^0  \\
\end{array}
\right) ,
\end{eqnarray}
where the unitarity matrices are given by
 \begin{eqnarray}
{\mathcal U_{L,R}}=\left( \begin{array}{cc}
c_{L,R} &  -s_{L,R} \\
s_{L,R} &  c_{L,R} \\
\end{array}
\right),
\end{eqnarray}
with $c_{L,R}=\cos(\theta_{L,R})$ and $s_{L,R}=\sin(\theta_{L,R})$. 
Thus the mass mixing matrix ${\mathcal M}$ is diagonalized as 
follows:

 \begin{eqnarray}
{\mathcal U_L}{\mathcal M}{\mathcal U_R^\dag}={\mathcal M_{diag}}=\left( \begin{array}{cc}
m_t &  0\\
0 &  m_T \\
\end{array}
\right),
\label{diag1}
\end{eqnarray}
or similarly
\begin{eqnarray}
{\mathcal U_{R,L}}\mathcal{M}^{\dag}\mathcal{M}{\mathcal U_{R,L}^\dag}=\mathcal{M}_{diag}^2 \, .
\label{diag2}
\end{eqnarray}
From the fact that the off-diagonal elements of ~(\ref{diag2}) vanish, one obtains the following relations for the mixing angles,
\begin{eqnarray}
\tan(2\theta_L)=\frac{4m_tm_T}{2m_T^2-2m_t^2-\lambda_T^2v^2}, \quad\quad
\tan(2\theta_R)=\frac{2\sqrt{2}\lambda_T m_tv}{2m_T^2+2m_t^2-\lambda_T^2v^2} .
\end{eqnarray}
We stress that the above mixing angles are not independent, 
by using eq ~(\ref{diag1}) one can derive the following relations
 \begin{eqnarray}
\tan\theta_R= \frac{m_t}{m_T} \tan\theta_L \quad , \quad
\frac{\lambda_T}{y_t} = \sin\theta_L \cos\theta_L \frac{m_T^2-m_t^2}{m_t m_T} \, .
\label{eq:lamT}
\end{eqnarray}
It is important to mention that both the interaction of the top-quark and of the T vector quark 
with the electroweak gauge bosons
depend on the mixing angle $\theta_L$ and are given by
\begin{eqnarray}
 \mathcal{L^{NC}} =\frac{g}{\cos\theta_W}Z_\mu&& \hspace*{-1.3em}\left(
\left[ \frac{\cos\theta_L^2 }{2} - \frac{2}{3} \sin^2\theta_W \right] \overline{t_L}\gamma^\mu t_L
+\left[ \frac{ \sin\theta_L^2}{2}  - \frac{2}{3} \sin^2\theta_W \right] \overline{T_L}\gamma^\mu T_L \right.
\nonumber \\ && \left.
+\; \frac{\sin\theta_L \cos\theta_L}{2} \left[ \overline{T_L} \gamma^\mu  t_L  + h.c\right]\right).
\end{eqnarray}
\begin{eqnarray}
 \mathcal{L^{CC}}=\frac{g}{\sqrt{2}}\left( \cos\theta_L \overline{t_L} +
 \sin\theta_L \overline{T_L} \right)\,b\gamma^\mu W_\mu^+ \;
 + \rm{h.c.} \, .
\end{eqnarray}

After EWSB one can derive the following Yukawa couplings of 
the CP-odd Higgs A to both $t$ and $T$
\begin{eqnarray}
At \bar t&=&\hspace*{-0.6em}-i\biggl(\frac{c_L c_R}{\tan\beta} -\frac{c_Ls_R y_T}{y_t}\biggr) \, , \nonumber \\
AT \bar T&=&\hspace*{-0.6em}-i\biggl(\frac{s_L s_R}{\tan\beta} +\frac{s_Lc_R y_T}{y_t}\biggr)  \, ,  \nonumber \\
A\bar t_L T_R&=&\hspace*{-0.6em}-i\biggl(\frac{c_L s_R}{\tan\beta} +\frac{c_Lc_R y_T}{y_t}\biggr)  \, ,   \nonumber \\
At_R\bar T_L &=&\hspace*{-0.6em}-i\biggl(\frac{s_L c_R}{\tan\beta}
-\frac{s_Ls_R y_T}{y_t}\biggr) \, .
\label{eq:rules}
\end{eqnarray}

After this brief review of the couplings of the heavy top to gauge bosons and
Higgs bosons, we list hereafter the most important theoretical and phenomenological 
constraints on the parameters of the model. From the theoretical side, the scalar sector of 2HDM+T is subject to the 
same unitarity and vacuum stability constraints as the usual 2HDM 
\cite{Branco:2011iw,Barroso:2013awa,Akeroyd:2000wc,Kanemura:1993hm}.
On the other hand, $y_T$ is also constrained from unitarity to be less 
than $4\pi$ while $\lambda_T$  is a derived quantity (see equation~\ref{eq:lamT}).

\begin{itemize}
\item The above interactions of $t_R$ and $T_R$ with the $Z$ and $W$ bosons 
in the 2HDM+T are the same as those  in the SM. 
Since the new top will contribute to gauge bosons self energies,
the  mixing angle $\theta_L$ can be constrained from electroweak observables 
such as S and T parameters \cite{Arhrib:2016rlj}. It has been shown in  
\cite{Arhrib:2016rlj} that $\sin\theta_L<0.2$ (resp 0.12) for $m_T=400$ GeV 
(resp $m_T=1$ TeV.)

\item The interaction of the heavy top with charged Higgs and bottom quark
can affect the rate of $BR(b\to s \gamma)$ \cite{Arhrib:2016rlj}. This can be
translated into constraint on the charged Higgs mass and/or mixing angle
between heavy top and top quark. In the case where we assume that
$\sin\theta_L \approx V_{Tb}$, it has been shown in \cite{Arhrib:2016rlj}  
that in 2HDM-II, $V_{Tb}$ must be smaller than $0.03$ for $m_T> 600$ GeV.
This limit can be weakened for light $m_T$.

\item Regarding compatibility with the couplings of the $h_{125}$ Higgs to the remaining
SM particles, since we are working in the alignment limit all couplings are SM-like
except the couplings to top quarks for which we choose a small mixing angle
to force compatibility with the measured Higgs couplings. Regarding the searches
for heavy scalars, the situation is exactly the same as for the 2HDM. So again
we use the limits derived for the scalar decay to fermions and study the effect
of the decay to gauge bosons in the parameter space of the model. 
\end{itemize}

Moreover, since the new heavy top couples to the all Higgs bosons, the decay
patterns $T\to bW, ht, Zt$ will be modified. It is well known that in the SM extended with a heavy top
the values of $Br(T\to bW)$, $Br(T\to ht)$ and $Br(T\to Zt)$ are
respectively 50\%, 25\% and 25\%. In the 2HDM with the presence of new Higgs bosons
$A, H$ and $H^\pm$, the above pattens are modified~\cite{Arhrib:2016rlj,Aguilar-Saavedra:2017giu}
and the limits obtained by ATLAS and CMS have to be reinterpreted in the framework of the 2HDM+T. \\
 
\section{Results for the 2HDM}
 
In this section we present the results for the $pp \to A \to VV$ production rates, where $V=\gamma, \, Z, \, W$, 
evaluated in the narrow width approximation. 
The pseudoscalar production cross section was calculated 
using SUSHI~\cite{Harlander:2012pb} at NNLO
and it includes gluon and $b \bar b$ fusion. 
The branching ratios were calculated using 
the HDECAY~\cite{Djouadi:1997yw, Djouadi:2006bz, Djouadi:2018xqq} version for 
the 2HDM~\cite{Harlander:2013qxa}.
The widths for the pseudoscalar decays into vector bosons are 
loop induced and were  calculated with the packages 
FeynArts~\cite{Hahn:2000kx}, FormCalc~\cite{Hahn:1998yk} and 
LoopTools~\cite{Hahn:1998yk,vanOldenborgh:1989wn} for loop integrals evaluation.
The loop calculation is performed in the 'tHooft-Feynman gauge using 
dimensional regularization. 
 At the one loop level, only fermionic loops contribute to 
$A \to VV$. The reason is that in the bosonic sector, the electroweak theory 
conserves CP while after adding fermions CP is no longer conserved. 
Therefore, there is no contribution to $A\to VV$ from the SM bosons 
\cite{Gunion:1991cw}, or from the spin zero scalars but only from the fermions.
 Analytical and numerical check of UV finiteness have been performed. 

As the measurements of the Higgs couplings to the other SM particles 
become increasingly precise, the 2HDM 
approaches more and more the alignment limit where 
$\sin (\beta -\alpha) \approx 1$ if $h$, the lightest CP-even scalar, is the SM-like Higgs boson.
Considering the lower limit on the charged Higgs mass (about 100 GeV
for Type I and X and 600 GeV for Type II and Y) and that 
$m_h = 125$  GeV, the decay $A \to h Z$ is kinematically  allowed
 for a pseudoscalar mass of more than about 200 GeV for all types while  $A \to H^\pm W^\mp$ would be allowed for a 
 pseudoscalar mass of about 200 (700) GeV  for Type I and X (Type II and Y).
In this study we will choose the heavy Higgs masses, $m_{H^\pm}$ and $m_H$, to be equal and such
that the decay $A \to H^\pm W^\mp$ is kinematically disallowed,
which is already the case for models Type II and Y due to the combined experimental 
and theoretical limits. The decay $A \to h Z$ is exactly zero in the alignment limit.
By extending this condition to all models we are being conservative in the chances
for the detection of a pseudoscalar in $VV$ final states. In fact, as more decays of the pseudoscalar 
are kinematically available, the chances of detecting a pseudoscalar in $VV$ final
states become smaller. 
The range of variation of the pseudoscalar mass is chosen to be in the range 
$50$ GeV $ < m_A < 600$ GeV for $A \to \gamma \gamma$ and  $2 \, m_V < m_A < 600$ GeV for $A \to VV$, $V=W,\, Z$.
In this mass range the main decay channels of the pseudoscalar are $A \to b \bar b$, $A \to \tau^+ \tau^-$ and $A \to \gamma \gamma$
(also $A \to ZZ (W^+ W^-)$ are possible but at much lower rate as will be discussed later). 
Clearly, the $A \to \gamma \gamma$ rate is at least two orders of magnitude below the tree-level decays
so one could ask how important this decay really is. 
Taking into account
the analysis performed for the 125 GeV SM Higgs where the ratio
\begin{eqnarray}
BR(h \to \gamma \gamma)/BR(h \to \tau^+ \tau^-)
\approx 0.0362
\label{eq:ratio}
\end{eqnarray}
holds and still the two photons channel was the first one to be measured,
we can expect that for a factor of about 100 the 
$A \to \gamma \gamma$ channel will 
still be competitive due to sharp resolution in the di-photon invariant mass achievable by the CMS and ATLAS detectors. 
The above number will be used a rough guide to what can be expected from future analyses. 

\begin{figure}[h!]\centering
 \includegraphics[width=0.5\textwidth]{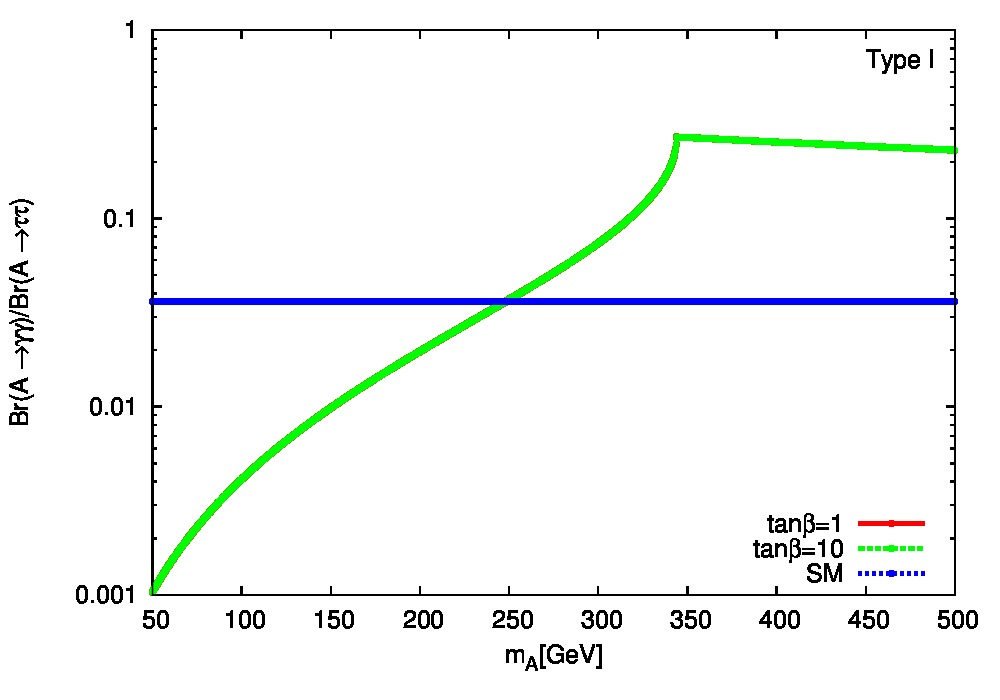}\includegraphics[width=0.5\textwidth]{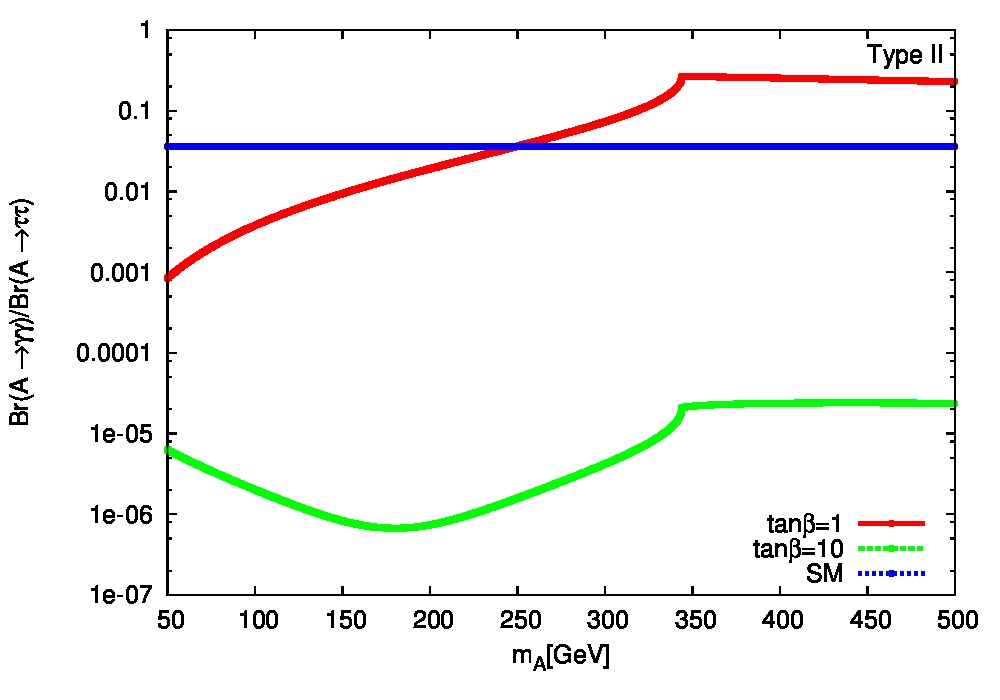}\\
\caption{$BR(A \to \gamma \gamma)/BR(A \to \tau \tau)$ as a function of $m_A$ for $\tan \beta =1$ and $\tan \beta =10$
in Type I and Type II.}
\label{comparison}
\end{figure} 

In figure~\ref{comparison} we show
$BR(A \to \gamma \gamma)/BR(A \to \tau \tau)$ as a function of $m_A$ for $\tan \beta =1$ and $\tan \beta =10$
in Type I and Type II. Also shown is the SM line for the same ratio. 
In Type I the two lines for $\tan \beta =1$ and $\tan\beta= 10$
 overlap because $A\to \gamma\gamma$ is mediated only through fermionic loops and the $Af\bar{f}\propto 1/\tan\beta$ coupling 
factorizes out and cancels with the same factor coming from $A \to \tau \tau$.
In Type I the SM line is crossed for 
$m_A \approx 250$ GeV independently of $\tan \beta$. However, since both the production cross section 
and the luminosity will be much higher during Run 2 it is to be expected that at very high luminosity 
all values of the pseudoscalar mass will be probed by the end of the LHC 14 TeV run in the low
$\tan \beta$ region.
The same behaviour is seen in Type II for low $\tan \beta$. As $\tan \beta$ increases it will become increasingly
harder to detect a pseudoscalar in the two-photons final state. 
It should be mentioned that the width of $A \to VV$, whatever the final state is, is controlled
by the top-quark loop. This loop is always proportional to $1/\tan^2 \beta$ and therefore the width
is the same for all Yukawa versions of the model except for very large values of $\tan \beta$
in Type II and Y. The behaviour we see in the right panel of figure~\ref{comparison} is due to
the width of $A \to \tau \tau$ that increases with $\tan \beta$ in Type II.  

\subsection{$A \to \gamma \gamma$} 

We now move to the detailed study of the production rates of a pseudoscalar decaying into two photons
at the LHC at 8 TeV and 14 TeV.
%
\begin{figure}[h!]\centering
 \includegraphics[width=0.5\textwidth]{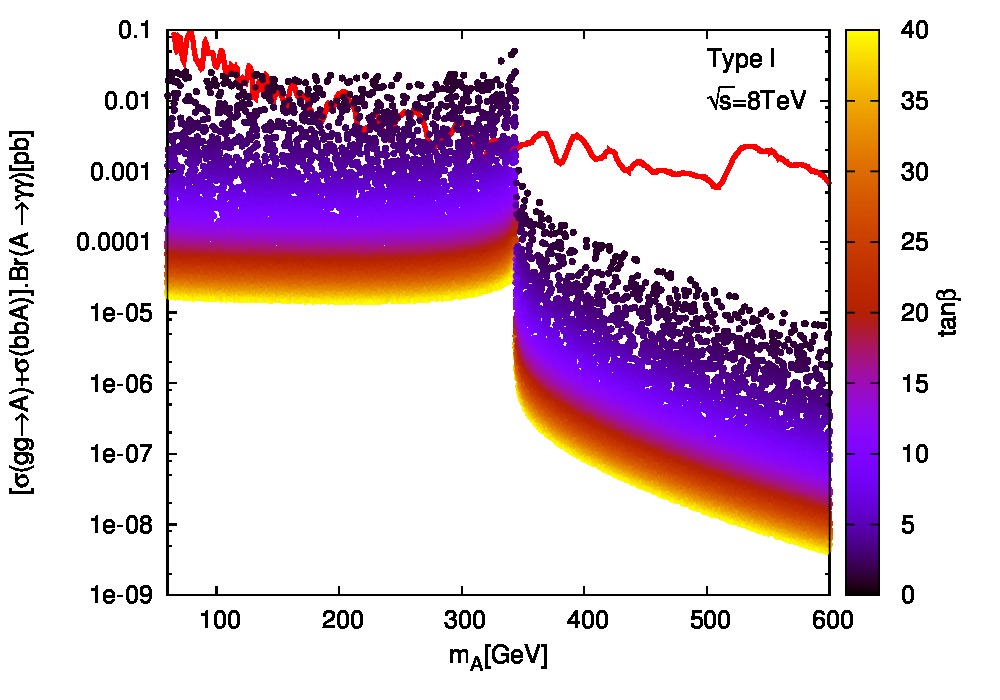}\includegraphics[width=0.5\textwidth]{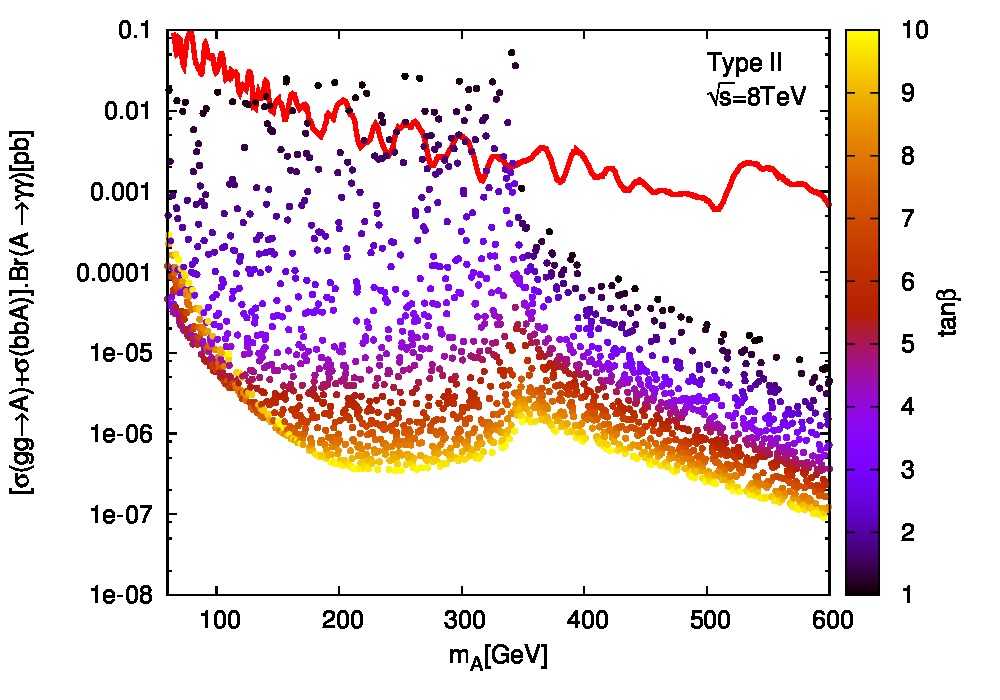}\\
 \includegraphics[width=0.5\textwidth]{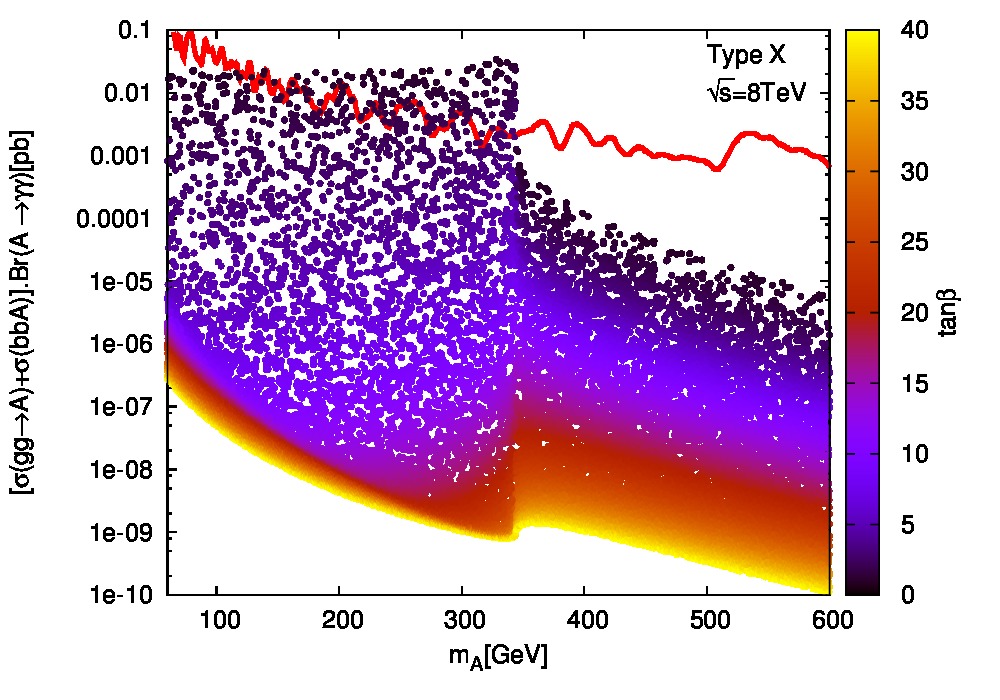}\includegraphics[width=0.5\textwidth]{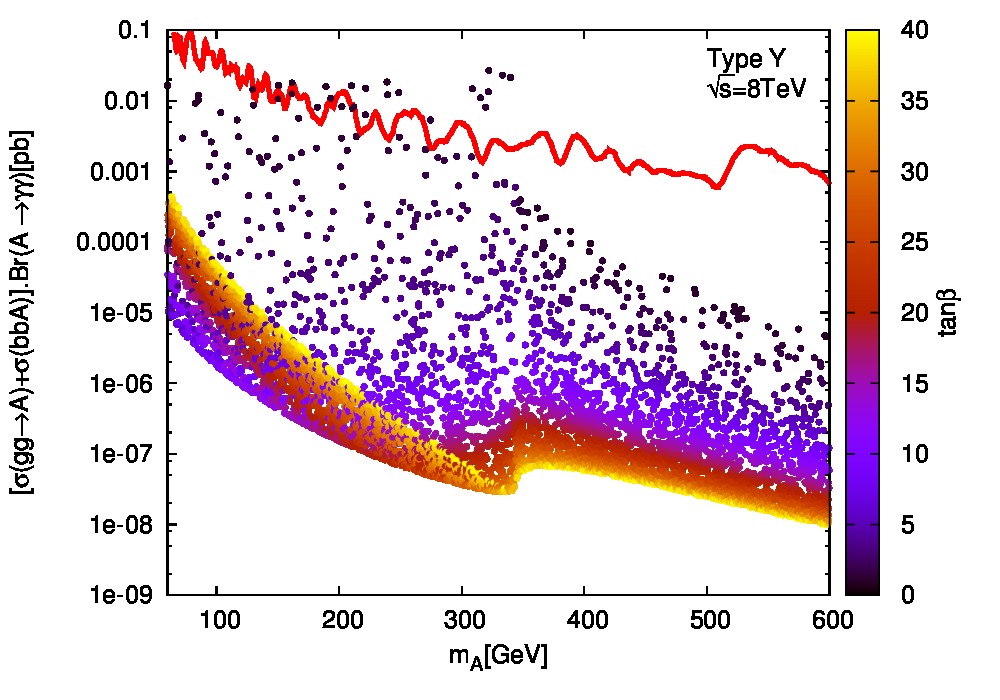}
\caption{$[\sigma(gg\rightarrow A)+\sigma(bbA)]BR(A\rightarrow \gamma \gamma)$ at $\sqrt{s}=8$ TeV  as a function of $m_A$ with $m_{H^+}=m_{H}=600 \, GeV$. The values of $\tan \beta$ are color coded
as indicated on the right of the plots. Also shown is the exclusion line from ATLAS (see text).}
\label{siggamma8}
\end{figure}
In figure~\ref{siggamma8} we plot
the pseudoscalar production cross section multiplied by the branching ratio 
$Br(A \to \gamma \gamma)$
for all four Yukawa types as a function of $m_A$. The remaining masses are fixed at $m_h = 125$ GeV 
and $m_{H^+}=m_{H}=600$ GeV.
Regarding the angles we take the exact alignment limit $\sin (\beta - \alpha) =1$ and we scan
over $\tan \beta$ from 1 to 40, except for Type II, where the scans stops at 10. As previously discussed,
the largest rates are for small $\tan \beta$ and therefore the upper bound is not relevant for the discussion. 
In the same plots we show the limits obtained by ATLAS~\cite{Aad:2014ioa} and 
CMS~\cite{Khachatryan:2015qba,CMS-PAS-HIG-14-037} after Run 1.
The situation is similar for all models: only a small region of the parameter space where $\tan \beta$
is small is excluded with this search. If we move to the large $\tan \beta$ and/or to large pseudoscalar
mass the number of events becomes negligible. The increase in cross section in Type II and Y due
to $bb$ production (the $bbA$ coupling is proportional to $\tan \beta$) is not enough to compensate
the decrease in branching ratio. Therefore the number of events will be small even for large $\tan \beta$
except for the region of very small pseudoscalar mass.

\begin{figure}[h!]\centering
 \includegraphics[width=0.5\textwidth]{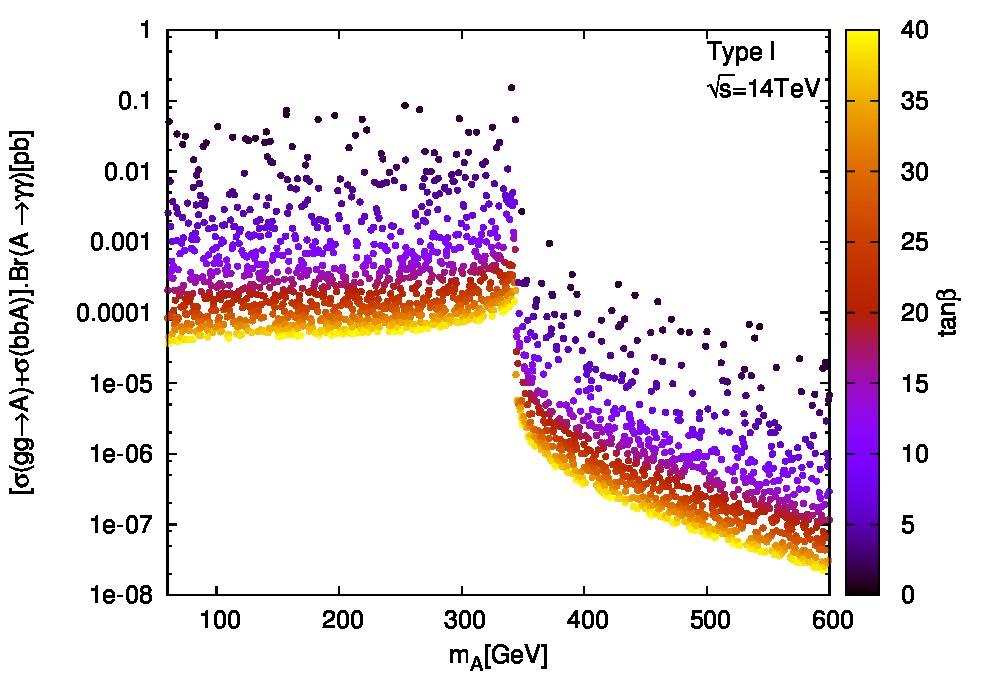}\includegraphics[width=0.5\textwidth]{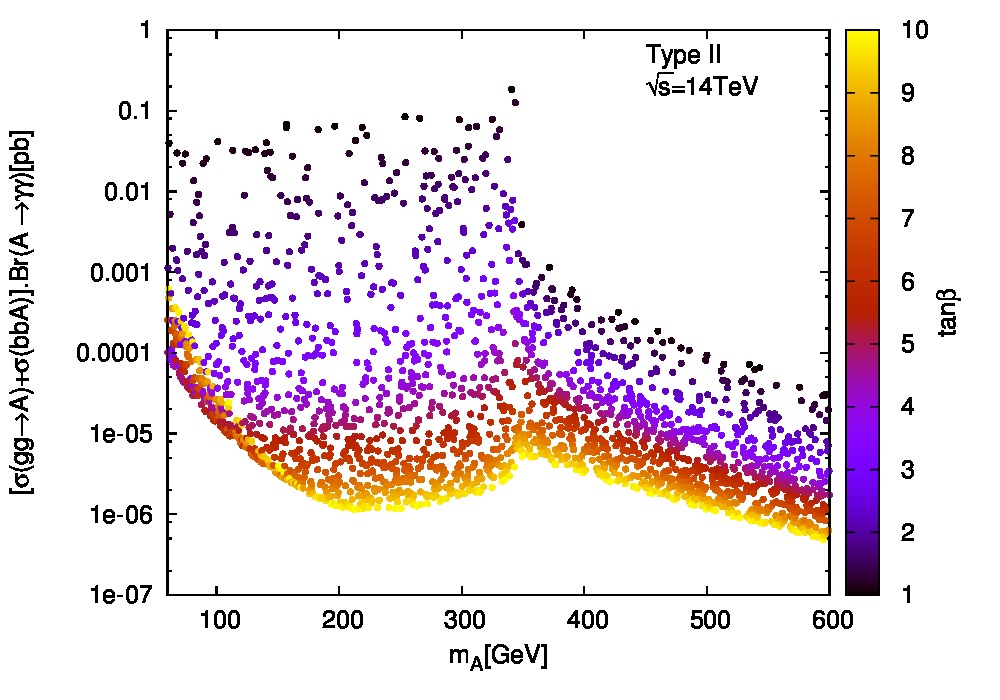}\\
\caption{$[\sigma(gg\rightarrow A)+\sigma(bbA)]BR(A\rightarrow \gamma \gamma)$ at $\sqrt{s}=14$ TeV as a function of $m_A$ with $m_{H^+}=m_{H}=600 \, GeV$. The values 
of $\tan \beta$ are color coded as indicated on the right of the plots.}
\label{siggamma14}
\end{figure}

In figure~\ref{siggamma14} we again show a scan in the Type I and Type II models 
for 14 TeV and with the exact same conditions of the previous figure~\ref{siggamma8}.
The pseudoscalar production cross section increases from a factor of about 2 
for $m_A = 50$ GeV to a factor of about 4 for $m_A = 600$ GeV. An extra
factor coming from the $bb$ initiated process will further increase the production
cross section for Type II and Y especially for pseudoscalar masses above
the $t \bar t$ threshold. Still this factor is quite small: taking for instance
$m_A = 500$ GeV the cross section increases by about 20 \% for $\tan \beta = 40$.  
Regarding the luminosity, which is expected to be about 300$fb^{-1}$ at the end of run 2, 
when compared to the 8 TeV run where the total luminosity collected was
about 30$fb^{-1}$, there is an approximate 10 fold increase. Overall a factor
below 20 can be foreseen for the low mass region. 
Therefore, it is clear that it will be hard to probe this channel above about $\tan \beta=10$ (and
probably less) by the end of Run 2. 

\subsection{$A \to ZZ (W^+W^-)$}  

 \begin{figure}[h!]\centering
  \includegraphics[width=0.5\textwidth]{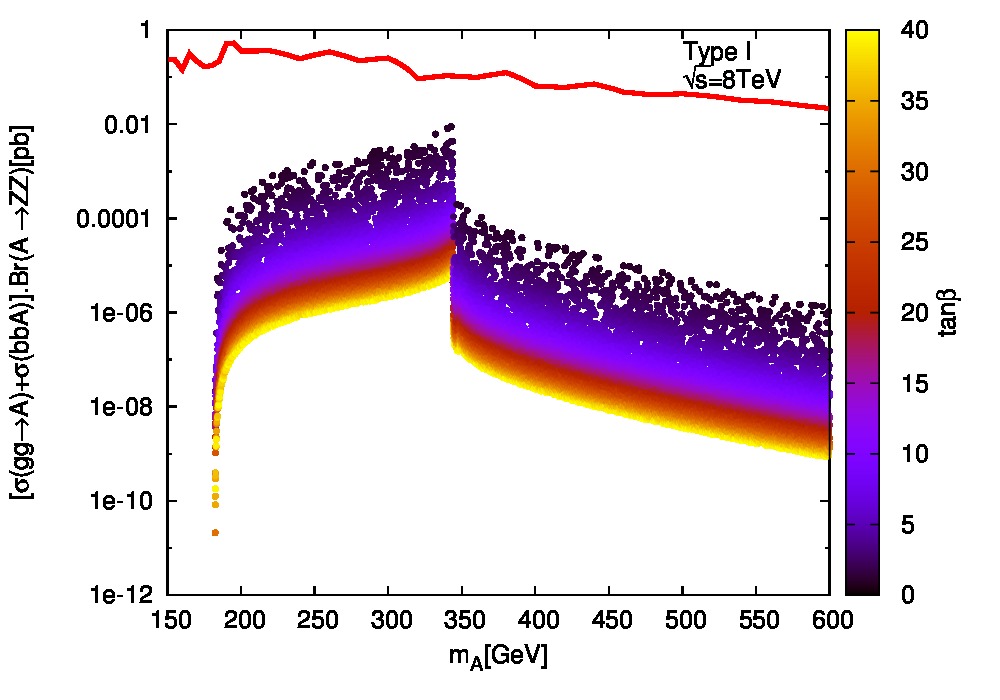}\includegraphics[width=0.5\textwidth]{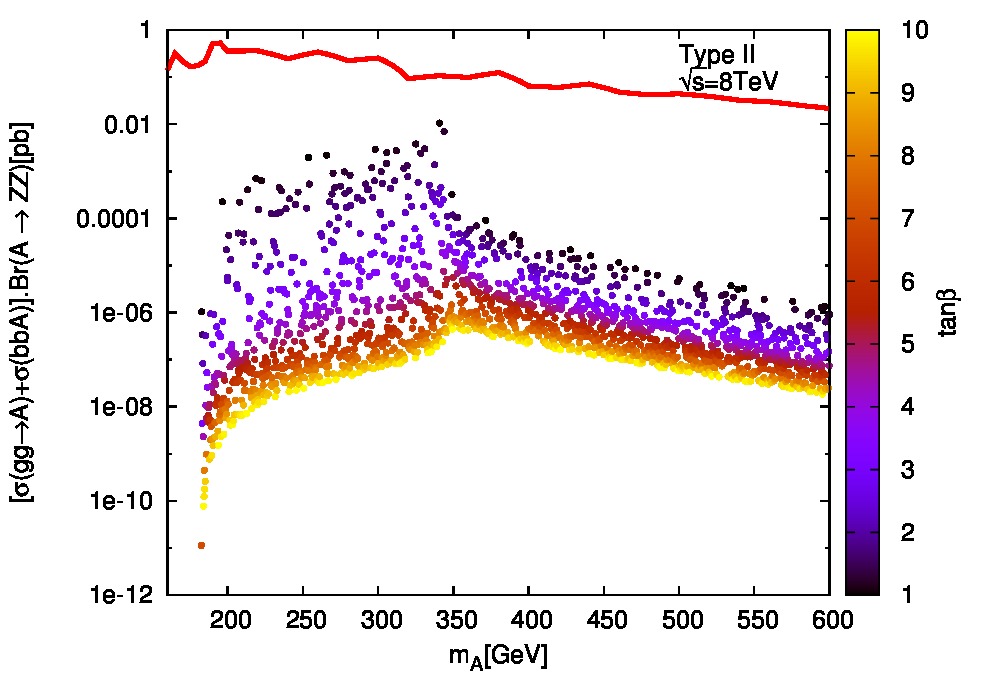}\\
  \includegraphics[width=0.5\textwidth]{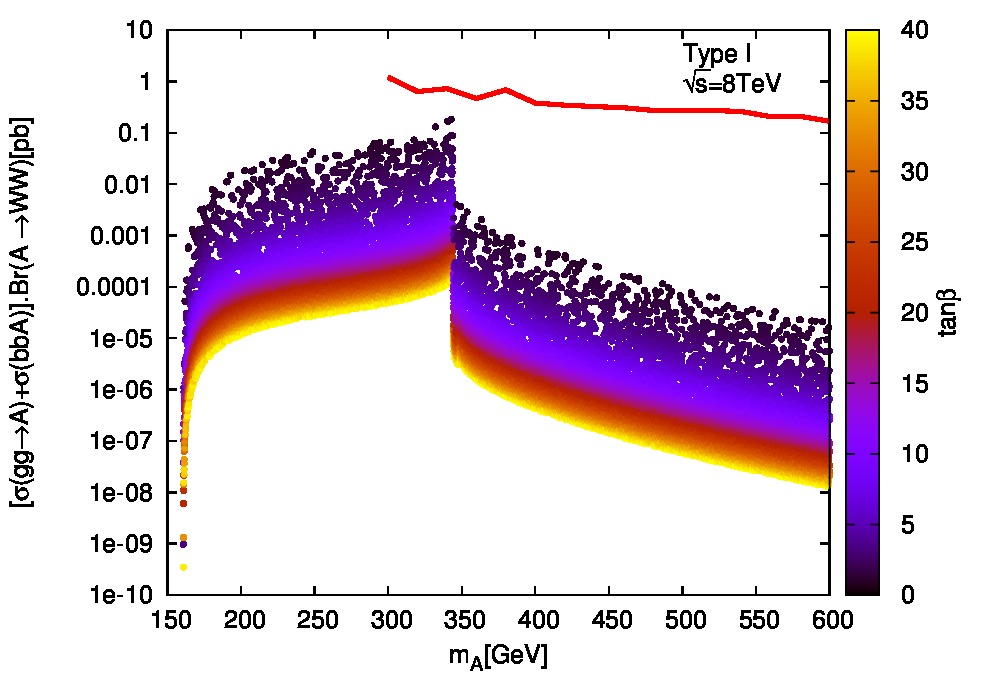}\includegraphics[width=0.5\textwidth]{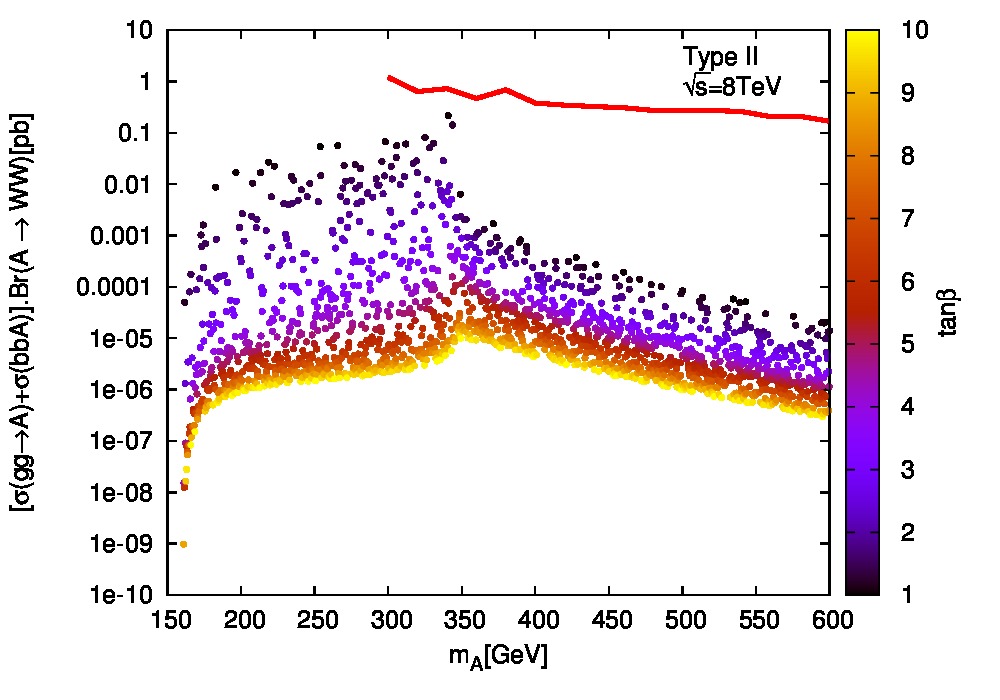}
 \caption{$[\sigma(gg\rightarrow A)+\sigma(bbA)]BR(A\rightarrow ZZ)$ (top)
 and $[\sigma(gg\rightarrow A)+\sigma(bbA)]BR(A\rightarrow WW)$ (bottom) at $\sqrt{s}=8$ TeV as a function of $m_A$
 with $m_{H^+}=m_{H}=600 \, GeV$. The values of $\tan \beta$ are color coded as indicated on the right of the plots.}
\label{sigmassive8}
 \end{figure}

\begin{figure}[h!]\centering
  \includegraphics[width=0.49\textwidth]{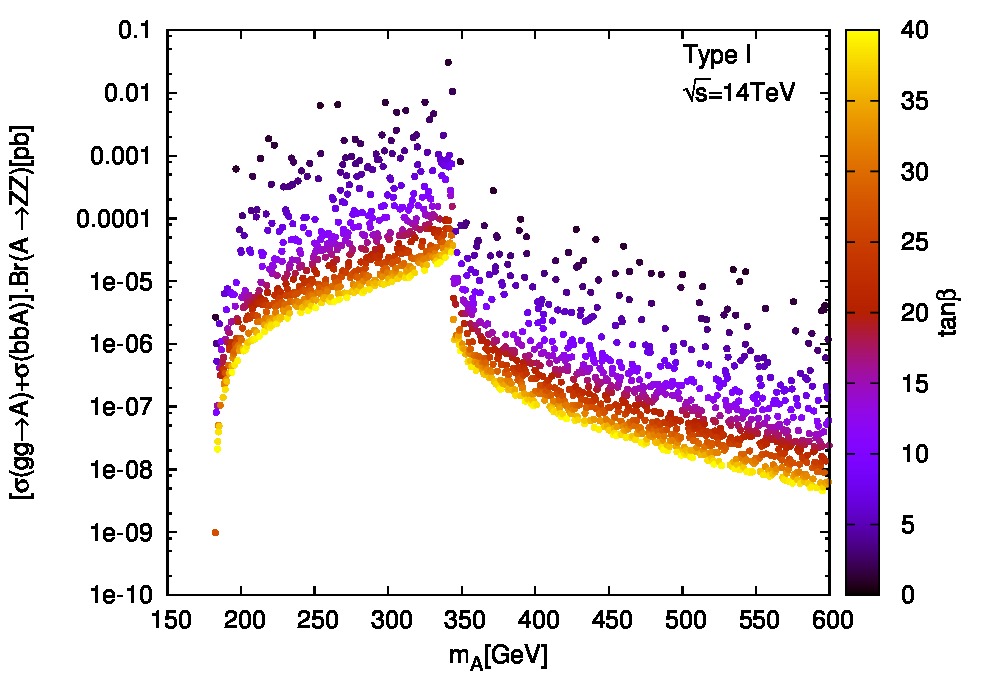}
  \includegraphics[width=0.49\textwidth]{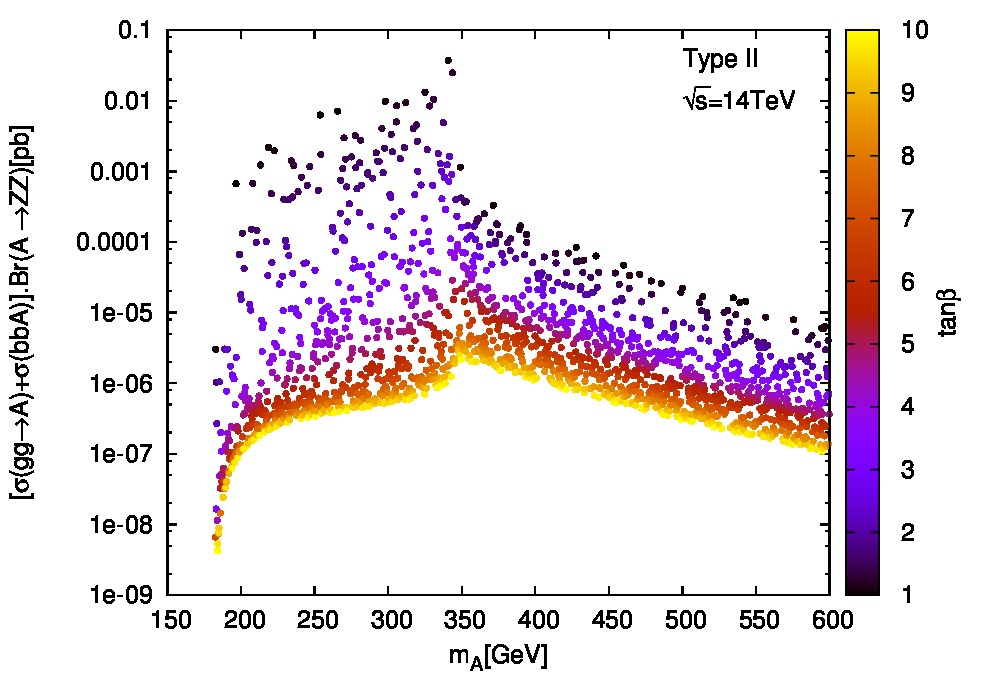}
 \caption{$[\sigma(gg\rightarrow A)+\sigma(bbA)]BR(A\rightarrow ZZ)$ at $\sqrt{s}=14$ TeV as a function of $m_A$
 with $m_{H^+}=m_{H}=600 \, GeV$. The values of $\tan \beta$ are color coded as indicated on the right of the plots.}
\label{sigmassive14}
 \end{figure}

The decay rate of a pseudoscalar to massive gauge bosons in the 2HDM is extremely small. The reason is clear:   
the decays of pseudoscalar bosons to massive gauge bosons can only occur at the loop level and
massive gauge bosons are heavy. In fact, $\Gamma (A \to ZZ)$
is always smaller than $10^{-5}$ GeV below the $t \bar t$ threshold and smaller than $10^{-4}$ GeV above
the same threshold, independently of the Yukawa version of the model. It is therefore clear
that these are not competitive channels when compared to the ones with two fermions or
even two photons final states. Since we are considering pseudoscalar production via fermion 
loops or $b \bar b$ initiated processes, a pseudoscalar decaying to two massive gauge
bosons is expected to be observed well after being detected in some fermion final state ($\tau^+ \tau^-$, $b \bar b$ or $t \bar t$)
or in $\gamma \gamma$. 
Similarly to the two-photon final state, also here we can use 
 exclusion bounds from 
searches  for a scalar decaying to either two $Z$ bosons or to $W^+ W^-$ performed by 
the ATLAS~\cite{Aad:2015agg, Aad:2015kna} and CMS~\cite{Khachatryan:2015cwa} collaborations. 
In figure~\ref{sigmassive8} 
we present the pseudoscalar production cross section multiplied by the branching ratio $A \to ZZ$ (top)
and  $A \to W^+W^-$ (bottom) for Type I (left) and Type II (right).
We also present the best experimental upper exclusion bound for these channels~\cite{Aad:2015agg, Aad:2015kna}. It is clear
that the experimental bounds are still about one order of magnitude away from
the points with the largest rates in the scan. Moreover, also for these final states only points below
the $t \bar t$ threshold and in the low $\tan \beta$ region have some chances of
being probed at the next LHC run. As previously discussed for the two photon final state
there is an overall factor of about 20 for the low mass region when considering both the increase in cross section and in luminosity. 
However, figure~\ref{sigmassive8}  clearly shows that even if the results are better
by two orders of magnitude we will barely start to probe a few scenarios in the 
low $\tan \beta$ region. In figure~\ref{sigmassive14} we present $[\sigma(gg\rightarrow A)+\sigma(bbA)]BR(A\rightarrow ZZ)$ at $\sqrt{s}=14$ TeV as a function of $m_A$
 with $m_{H^+}=m_{H}=600 \, GeV$.  As discussed, both for Type I and Type II, there is an increase in the maximum values of the rates but still well below the experimental
 result line and an increase in more than one order of magnitude is needed to start probing the largest values of the rates.

\section{Results for the 2HDM+T}


\begin{figure}[h!]\centering
 \includegraphics[width=0.45\textwidth]{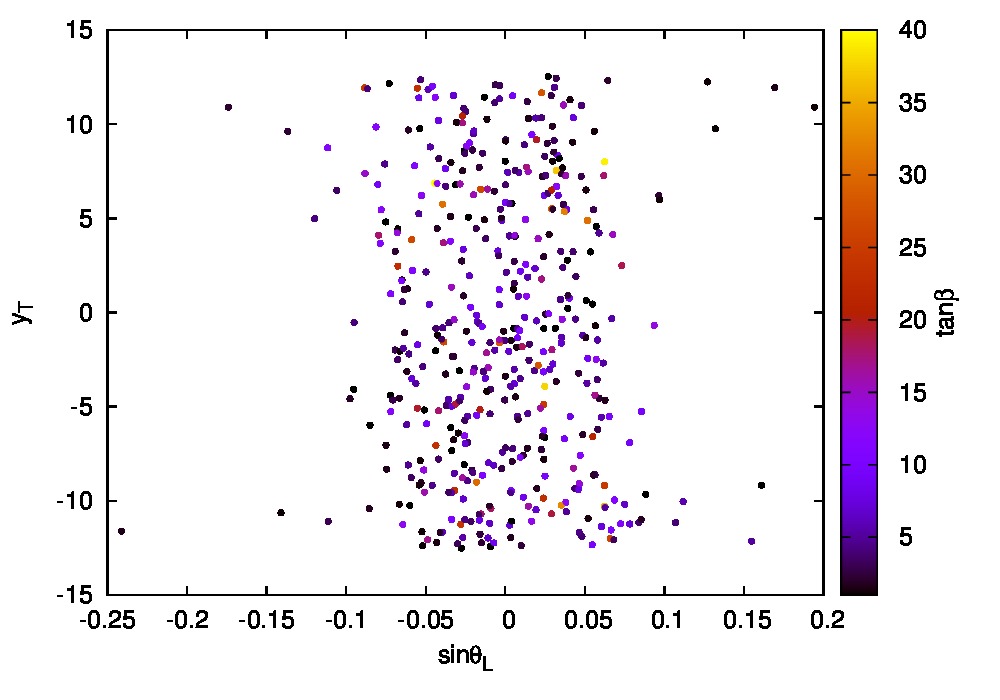}
\includegraphics[width=0.45\textwidth]{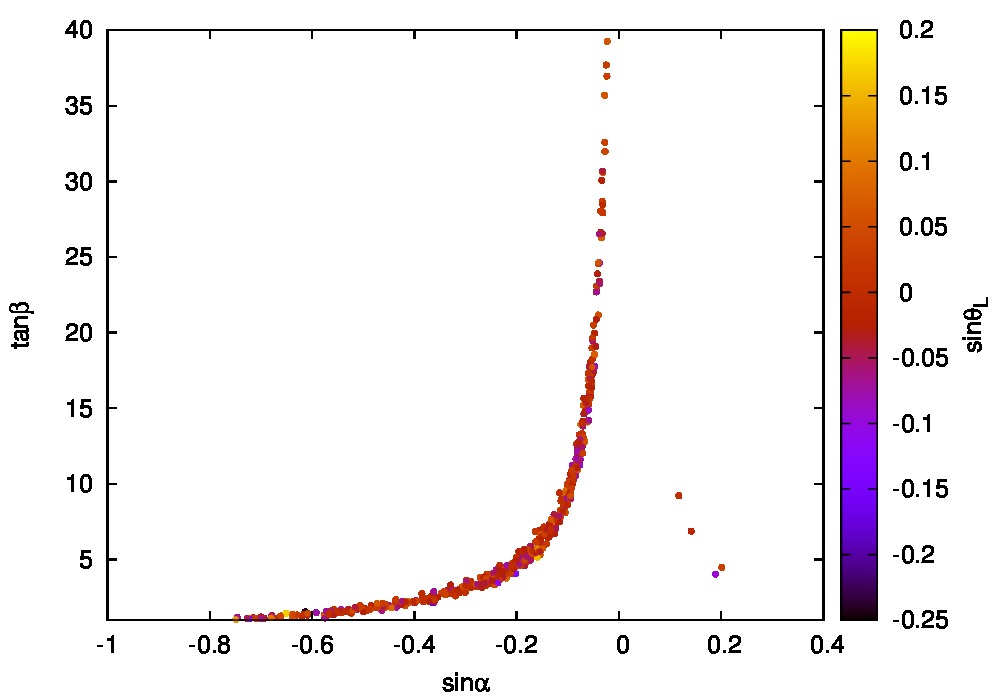}
\caption{ Results of the scan described in the text in the ($\sin\theta_L, y_L$) plane (left) and in the ($\sin\alpha, \tan\beta$) plane (right), after imposing the 
most relevant theoretical and experimental constraints as previously described in detail ($m_T=1$ TeV).}
\label{fig:feynman}
\end{figure}

\begin{figure}[h!]\centering
 \includegraphics[width=0.45\textwidth]{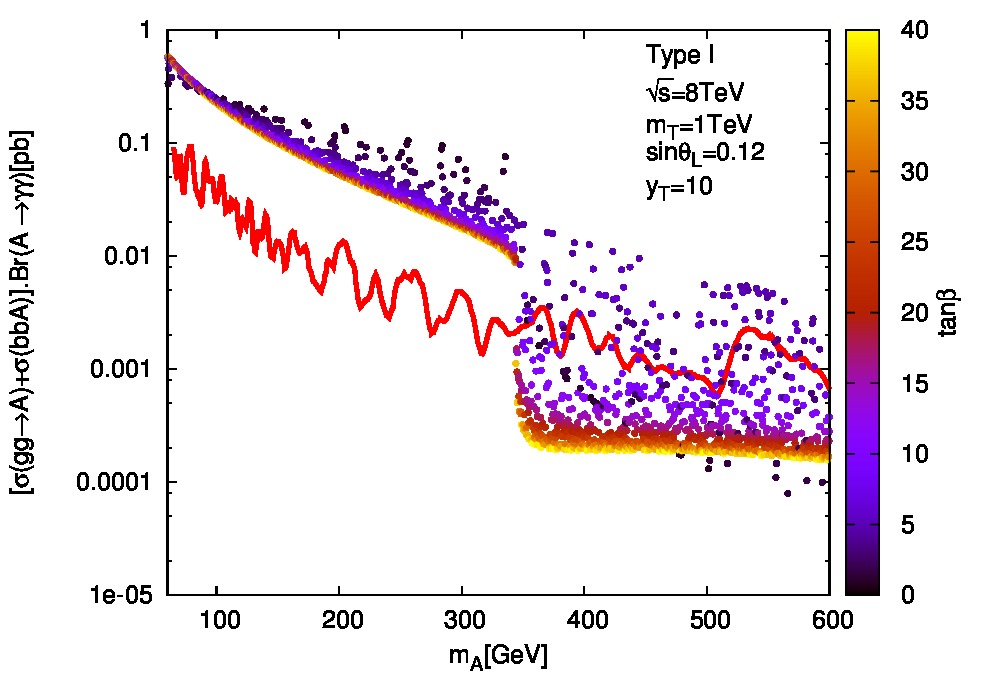}\includegraphics[width=0.45\textwidth]{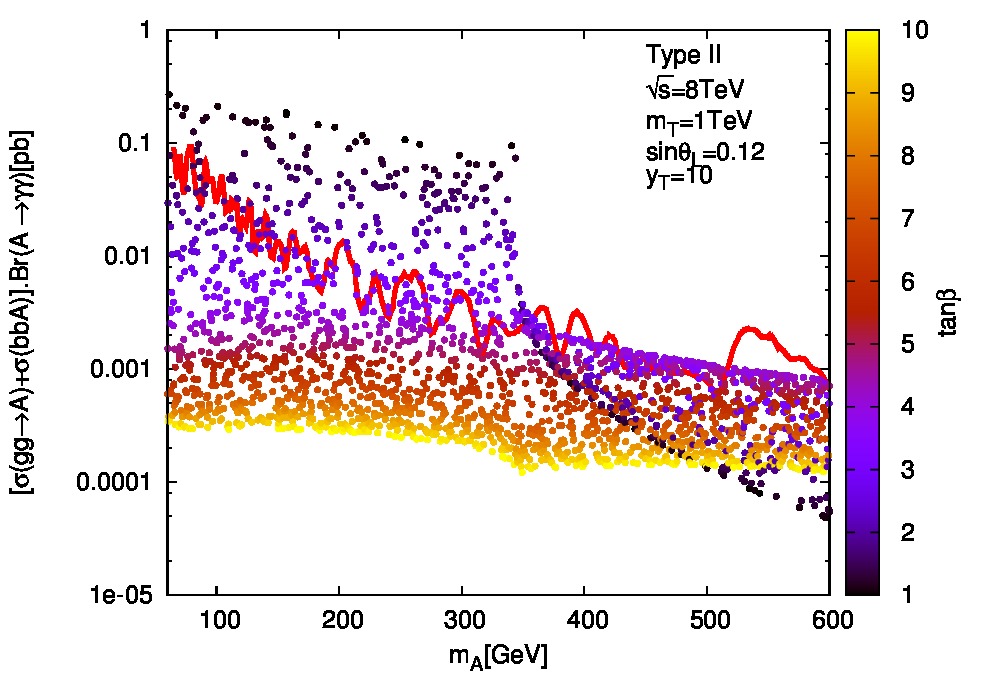}\\
 \includegraphics[width=0.45\textwidth]{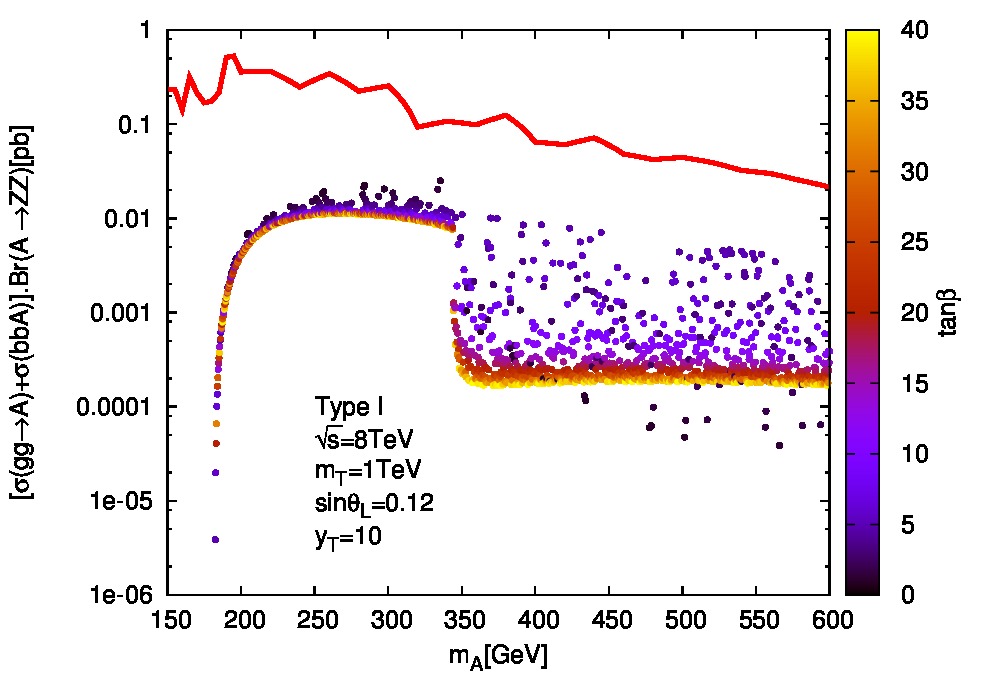}\includegraphics[width=0.45\textwidth]{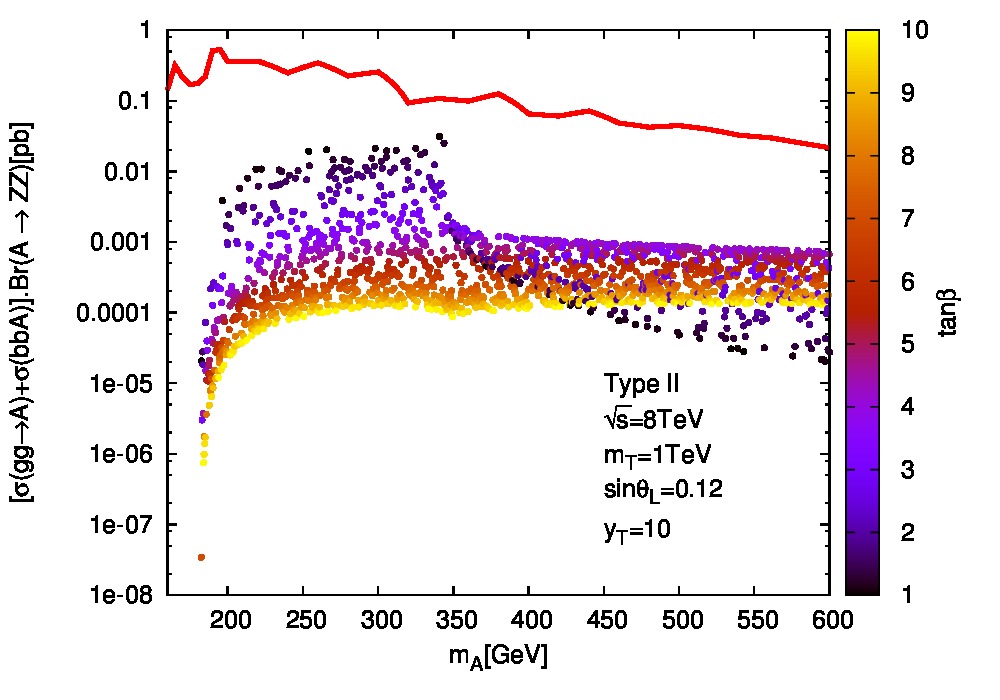}\\
  \includegraphics[width=0.45\textwidth]{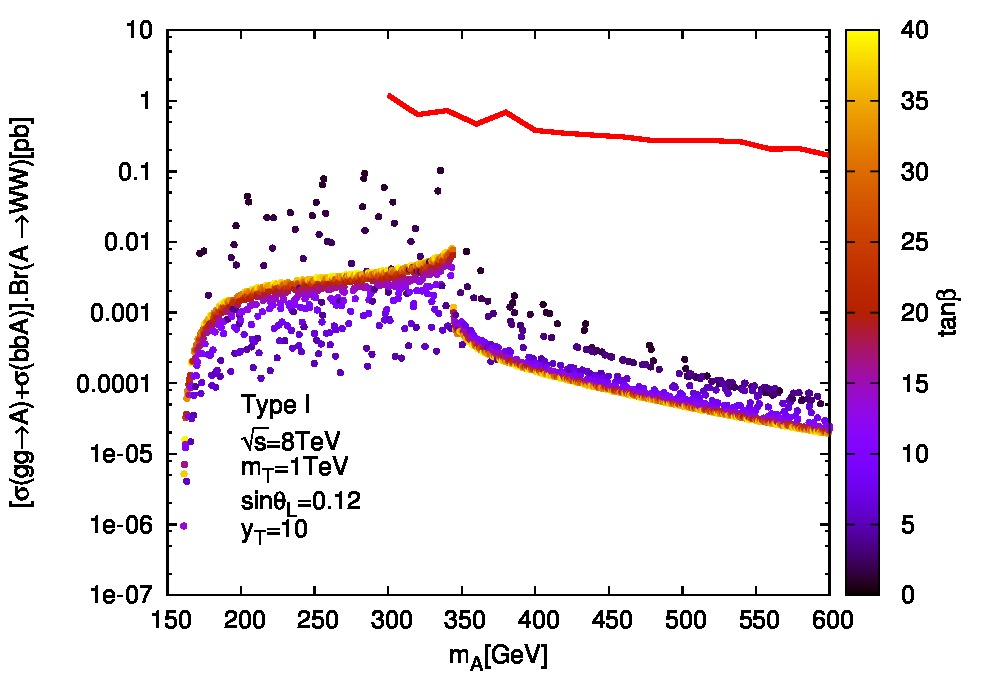}\includegraphics[width=0.45\textwidth]{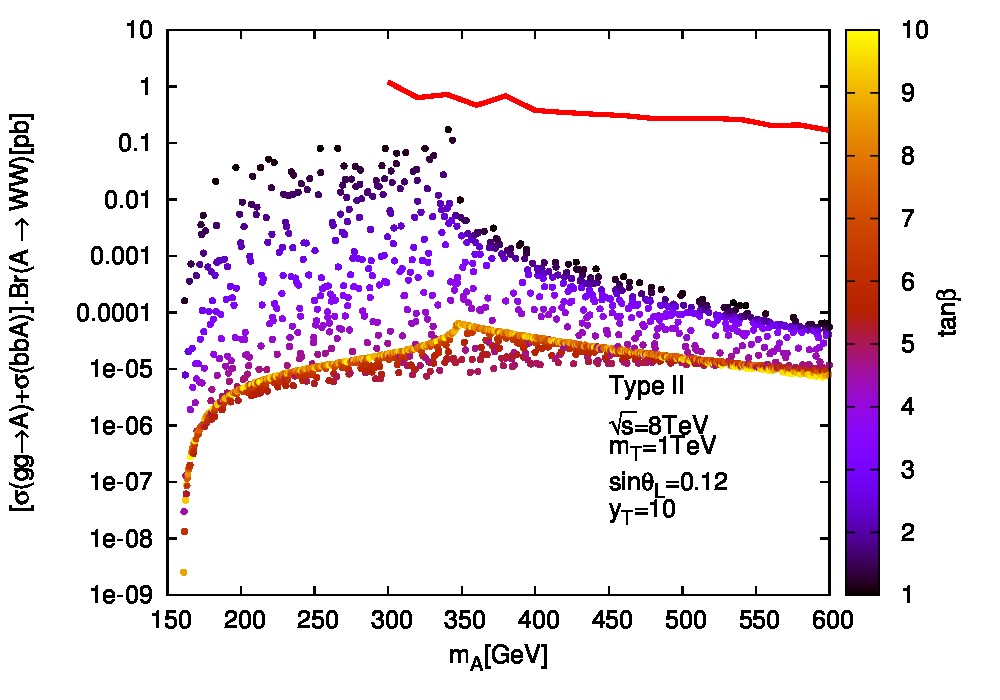}
\caption{Scatter plots of $[\sigma(gg\rightarrow A)+\sigma(bbA)] \, Br(A\rightarrow \gamma \gamma,ZZ,WW)$ at $\sqrt{s}=8$ TeV in the 2HDM+T as a function of $m_A$ where $m_{H^+}=m_{H^0}=600$ GeV, $\sin(\theta_L)$=0.12, $m_T=1$TeV and $y_T=10$,the values 
of tan $\beta$ are color coded as indicated on the right of the plots}
\label{2hdm-vlq-8tev}
\end{figure}


Before we present our results for 2HDM with a vector like top, we first show in figure~\ref{fig:feynman}
the allowed range for $y_T$, $\tan\beta$, the mixing angles $\alpha$ and $\sin\theta_L$ for a fixed $m_T=1$ TeV.
From the left plot of figure~\ref{fig:feynman} one can see that $|y_T|\leq 15$ is allowed and that the mixing angle $\sin\theta_L$ should be less 
than about 0.2 for any value of $1\leq \tan\beta\leq 40$. In the right panel we show $\tan \beta$ as a function of $\sin \alpha$ where
we can see a very similar behaviour to the 2HDM for the same variables, indicating that we are again very close to alignment
except for a few points in the region where $\sin \alpha > 0$ where the coupling to b-quarks change sign, known
as the wrong sign limit~\cite{Ferreira:2014naa, Ferreira:2014dya}.

Extending the 2HDM to include a vector-like top quark, results in an enhancement both in the pseudoscalar
production cross section and in its decay width into $\gamma \gamma$, $ZZ$ and $W^+W^-$. 
Hence, even with all constraints taken into account, in the case of large mixing 
between the top quark and new top T, the production rates of loop-induced 
processes can be several orders of magnitude above the 2HDM ones. 
In order to see the effects of the new top in the production rates
of the gauge bosons we plot in figure~\ref{2hdm-vlq-8tev} the rate
$[\sigma(gg\rightarrow A)+\sigma(bbA)] \, Br(A\rightarrow \gamma \gamma,ZZ,WW)$
at $\sqrt{s}=8$ TeV as a function of $m_A$ (and $\tan \beta$), where we
chose $m_T=1$ TeV, $y_T=10$ and $\sin(\theta_L)=0.12$.
One can see from this plot that the effect of the new top on 
the production and the decay of the CP-odd Higgs is a significant enhancement 
in both the production cross section by gluon fusion and in the decays
$A \rightarrow \gamma \gamma$, $A \to ZZ$ and $A \to W^+W^-$ when compared to the 
the ones in the 2HDM (note that the Higgs production in bottom-quark annihilation 
is not altered by the new top).

As it is clear from figure~\ref{2hdm-vlq-8tev}, the largest enhancement in the
rates is again obtained for low $\tan\beta$, as in the 2HDM. Like in the 2HDM, this behaviour can be understood
by looking at the couplings given in equation~(\ref{eq:rules}), valid both for 2HDM+T Types I and II, 
which show that all couplings contain a term proportional to $1/\tan\beta$. Another source of
enhancement relative to the 2HDM is the choice of large Yukawa $y_T$, which is still well below the perturbativity limit of $4\pi$, 
as well as large mixing  $\sin\theta_L$. Also shown in the plots are the exclusion lines from 
the experiments at CERN. In the upper plots we can see the diphoton exclusion line from ATLAS. 
This plot is presented to show a very interesting result:  in a Type I 2HDM+T a pseudoscalar with $m_A<2m_t\approx 350$ GeV and
large mixing $\sin\theta_L\approx 0.12-0.06$ is excluded for any value of $\tan\beta$. This means that the two-photon
final state search is important to further constrain models with vector like quarks. One should note however 
that away from $t\bar{t}$ threshold the exclusion is valid only for rather small $\tan\beta$. In the Type II 2HDM+T, 
one can see that for $m_A\leq 350$ GeV small $\tan\beta$ is excluded from diphoton events. We note that there is no exclusion for small 
$\sin\theta_L\leq 0.01$ both for Type I and Type II. In the middle and lower panels of figure ~\ref{2hdm-vlq-8tev} we present
  the rates for $\sigma(pp\to A)\times Br(A\to VV)$ for V=Z or W together with the exclusion line from ATLAS. In both cases, the total rates are still 
 about one order of magnitude smaller than the exclusion line. We note that there is a slight enhancement
 in the rate $\sigma(pp\to A)\times Br(A\to ZZ)$ with respect to the 2HDM case, due to the extra loop contribution.
 Still depending on the parameters chosen, the rate $\sigma(pp\to A)\times Br(A\to VV)$ for $V=Z, W$ may 
also be suppressed compared to the 2HDM.
 

\section{At a 100 TeV pp collider} 

In the quest for new physics, there is a consensus among the community of particle physicists
in favour of the construction a high energy machine with 100 TeV center of mass energy.
One question that has been raised about a future 100 TeV $pp$ collider is what is the luminosity
needed to address the physics that is not within the reach of the LHC, even at high luminosity.
In~\cite{Hinchliffe:2015qma} several physics scenarios were analysed and a luminosity of about
10-20 ab$^{-1}$ was shown to be a good compromise in extending the discovery reach for new phenomena 
relative to the high luminosity LHC.

 \begin{figure}[h!]\centering
  \includegraphics[width=0.5\textwidth]{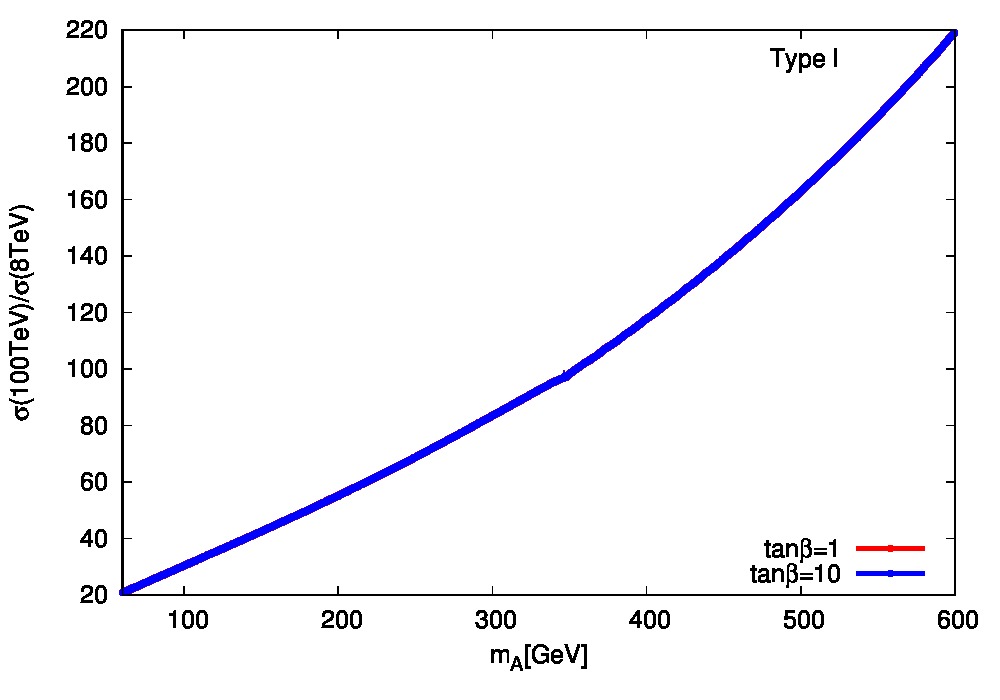}\includegraphics[width=0.5\textwidth]{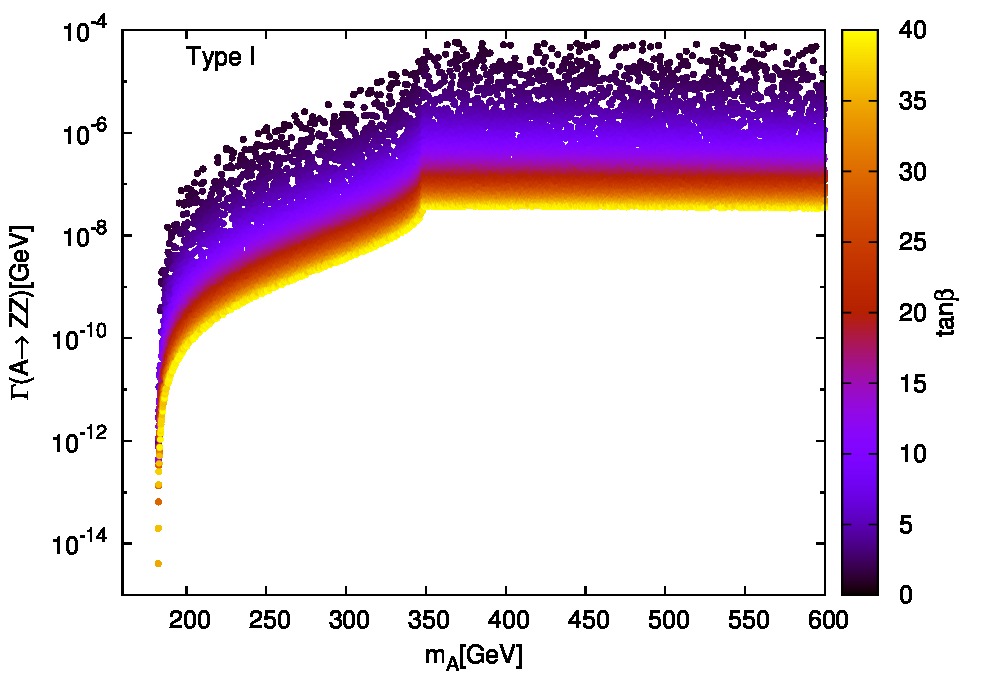}\\
 \caption{Left panel: ratio of cross section $\sigma(pp \to A)_{100 \, TeV}/\sigma(pp \to A)_{8 \, TeV}$
 as a function of $m_A$ for $\tan \beta = 1$ and $\tan \beta = 10$; 
 Right panel: partial width $\Gamma (A \to ZZ)$ as a function of $m_A$ for Type I.
 The values of $\tan \beta$ are color coded as indicated on the right of the plot. }
\label{2hdmt100tev}
 \end{figure}

The production cross sections $pp \to A$ at a 100 TeV collider is increased relative to the 
8 TeV LHC from a factor of about 20 for $m_A = 50$ GeV to about 220 for $m_A = 600$ GeV.
This behaviour is shown in the left panel of figure~\ref{2hdmt100tev} where the ratio
of the cross sections for 100 TeV and for 8 TeV $\sigma(pp \to A)_{100 \, TeV}/\sigma(pp \to A)_{8 \, TeV}$
is shown as a function of the pseudoscalar mass for Type I and two values of $\tan \beta$, 1 and 10.
The plots for all other Yukawa types show exactly the same behaviour as the one for Type I for $\tan \beta=1$
and for large $\tan \beta$ the contribution of the $bb$ initiated process slightly changes
this ratio with no meaningful changes in the conclusions.

As previously discussed, so far analyses were only performed for 8 TeV with a total luminosity of 
about 30$fb^{-1}$. Therefore, in the low mass region the cross section is increased by a factor of 20
while the luminosity is incremented by about 1000. Overall, an improvement of at least
four orders of magnitude is expected. The increase is more significant for higher masses
but the branching ratios are smaller. Furthermore, the ratio of the cross sections
is almost independent of $\tan \beta$ and of the Yukawa type. Considering figure~\ref{siggamma8} it is clear that
most of the parameter space will be probed in the case of $pp \to A \to \gamma \gamma$ for the 2HDM, and for any Yukawa type. However,
when examining figure~\ref{sigmassive8} for the case of the decays into massive gauge bosons,
we see that only a small portion of the parameter space will be probed, mainly for low $\tan \beta$
and for pseudoscalar mass below the $t \bar t$ threshold. The same is true for the $W^+ W^-$ final state.


\begin{center}
\includegraphics[width=0.45\textwidth]{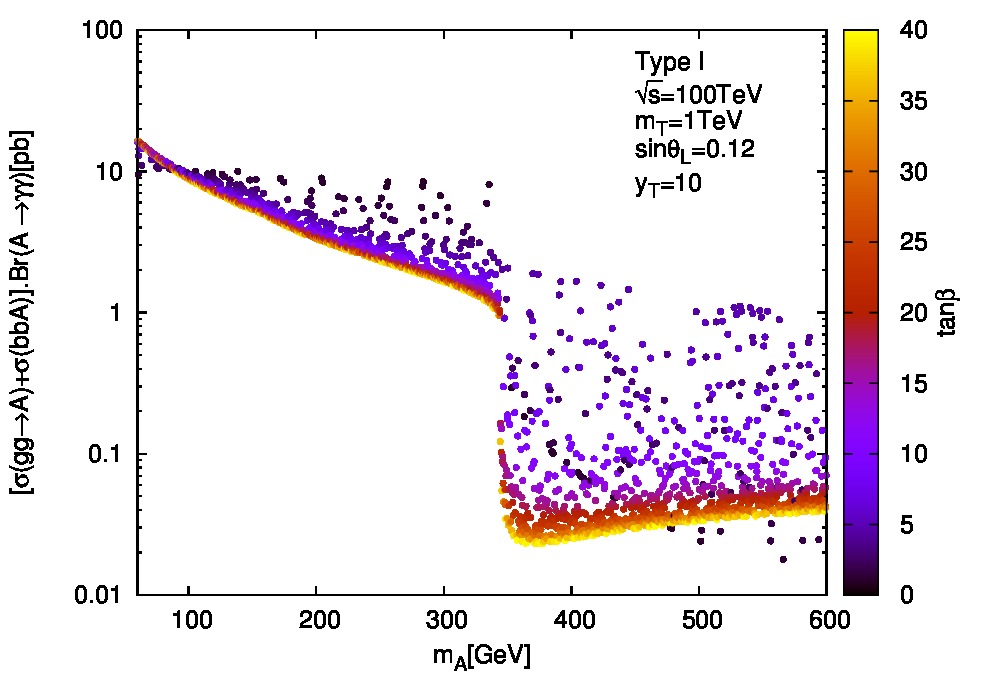}\includegraphics[width=0.45\textwidth]{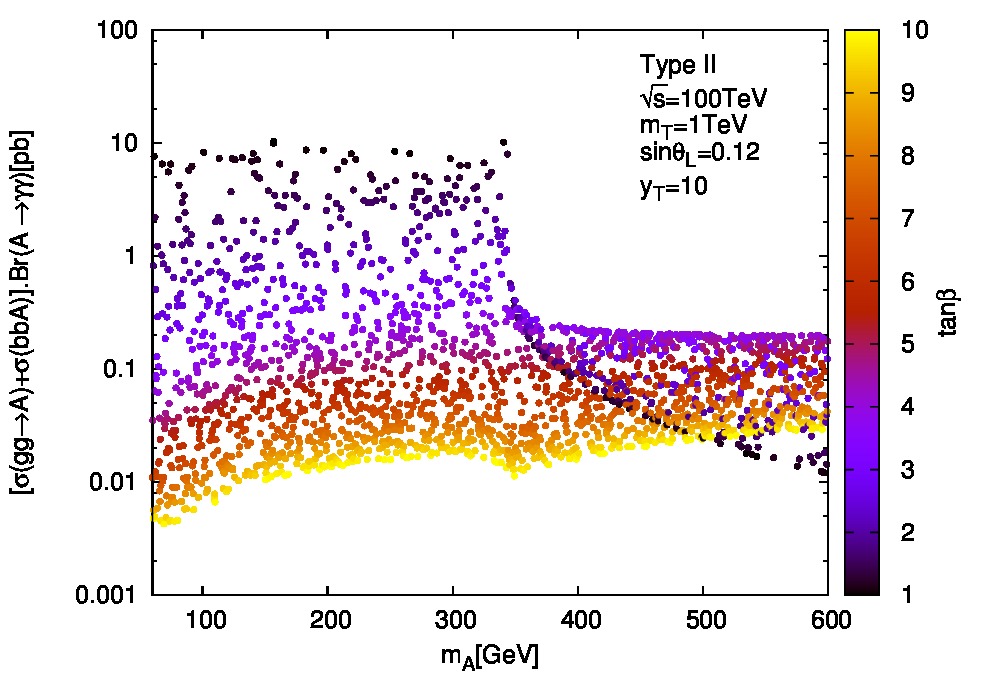}\\
 \includegraphics[width=0.45\textwidth]{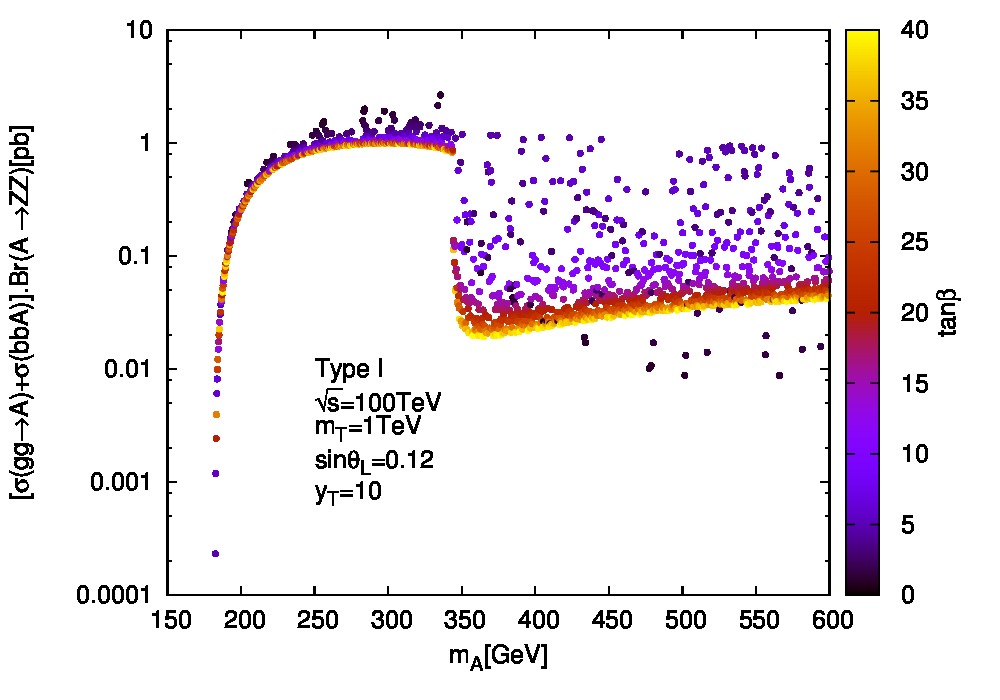}\includegraphics[width=0.45\textwidth]{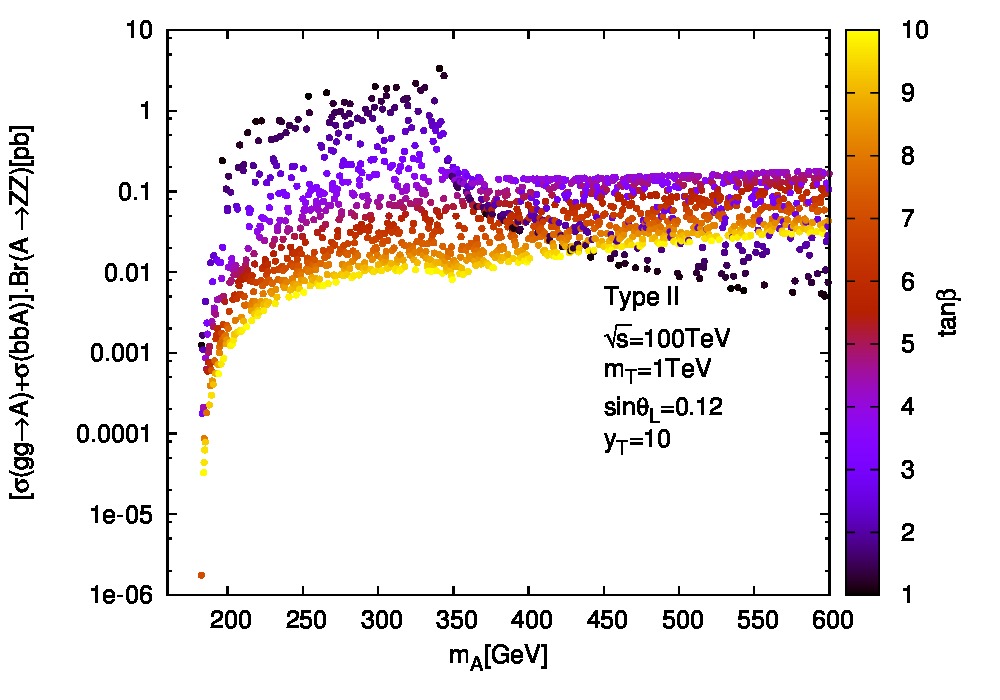}\\
  \includegraphics[width=0.45\textwidth]{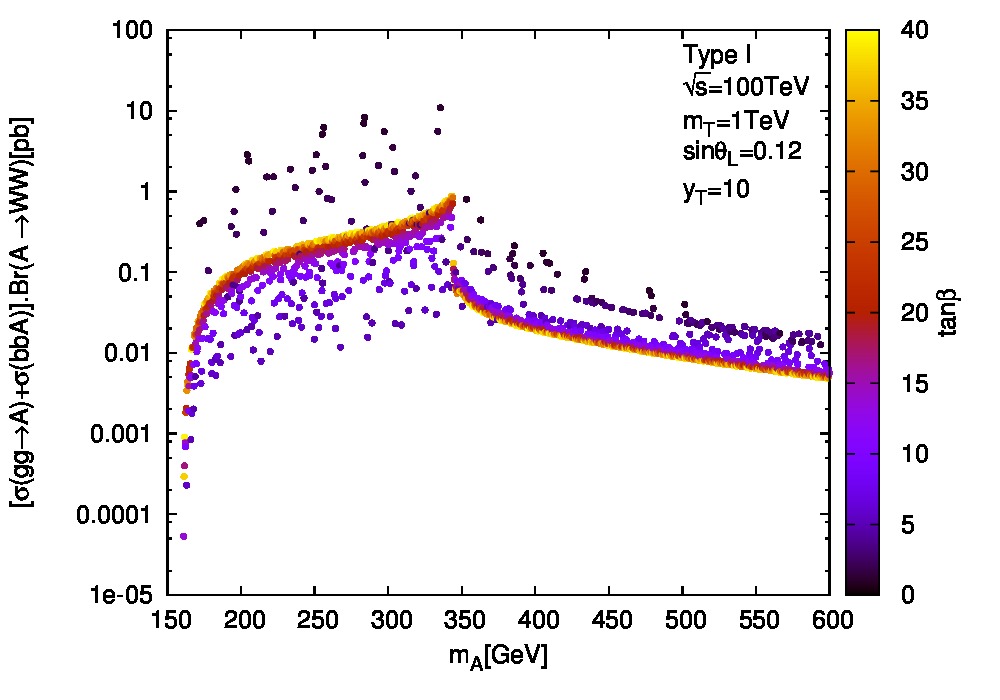}\includegraphics[width=0.45\textwidth]{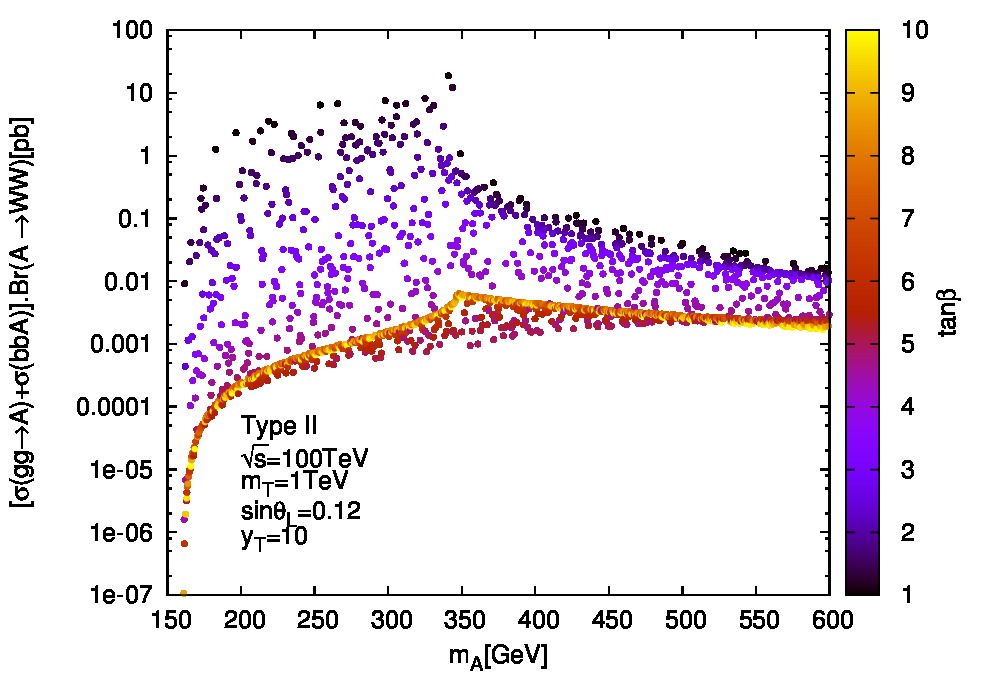}
\captionof{figure}{Scatter plots for $[\sigma(gg\rightarrow A)+\sigma(bbA)] \, Br(A\rightarrow \gamma \gamma,ZZ,WW)$ for $\sqrt{s}=100$ TeV in the 2HDM+T as a function of $m_A$ where $m_{H^+}=m_{H^0}=600$ GeV, $\sin(\theta_L)$=0.12,
$m_T=1$TeV and $y_T=10$,the values 
of tan $\beta$ are color coded
as indicated on the right of the plots.}
\label{2hdmplust100tev}
 \end{center}

Finally in figure~\ref{2hdmplust100tev} we present scatter plots for $[\sigma(gg\rightarrow A)+\sigma(bbA)] \, Br(A\rightarrow \gamma \gamma,ZZ,WW)$ for 
$\sqrt{s}=100$ TeV in the 2HDM+T as a function of $m_A$ where $m_{H^+}=m_{H^0}=600$ GeV, $\sin(\theta_L)$=0.12, $m_T=1$TeV and $y_T=10$. The left
plots are for Type I and the right plots are for Type II. In the case of $A \to \gamma \gamma$ part of the parameter space, for the low mass region is already
excluded for the parameters shown. For the same set of parameters almost all values of $m_A$ and $\tan \beta$ are within the reach of a 100 TeV colliders.
However, as $y_T$ decreases, the model will resemble the 2HDM and therefore as previously discussed only the low $\tan \beta$ region will have some chances
to be probed.

In order to roughly quantify the sensitivity for the $ZZ$ and $W^+W^-$ final states (where our main interest is focused), we will perform some rough estimates regarding the observability
of the CP-odd scalar decaying to two Z-bosons in the four-lepton channel. Let us start by computing
an upper limit of the number of events obtained for the 2HDM (by choosing $\tan \beta =1$) taking into account the efficiency 
factor for the four lepton channel search. As discussed in~\cite{Gunion:1991cw}, for the $4l$ channel, the only significant background is $qq, gg\to ZZ\to 4l$, 
which is fully determined by the detector resolution in the 4l channel \cite{Gunion:1991ps}.
For the efficiency factor for this channel we use the number given in \cite{Gunion:1991ps}, which is $\epsilon \approx 0.35$.
The number of 4l events that one would then obtain at a hadron collider is calculated as follows
\begin{eqnarray}
N_{events} &=& \sigma(pp\to A^0)\times Br(A^0\to ZZ) \times (Br(Z\to 2l))^2 \times {{\cal L}} \times \epsilon
\label{rawevents}
\end{eqnarray}
where we will use the predicted collected luminosity at the end of the 14 TeV LHC Run, ${{\cal L}}=3000 fb^{-1}$ ,
and the estimate collected luminosity for the 100 TeV machine of ${{\cal L}}=10 ab^{-1}$. This number of events
is then to be compared to the minimum number of events required to obtain 4$\sigma$ statistical significance as
calculated in~\cite{Gunion:1991cw}. In reference~\cite{Gunion:1991cw} this significance was estimated to lead to about 
20 events for $m_A\approx 200$ GeV and 10 events for $m_A\approx 400$ GeV. 

\begin{figure}[h!]\centering
\includegraphics[width=0.45\textwidth]{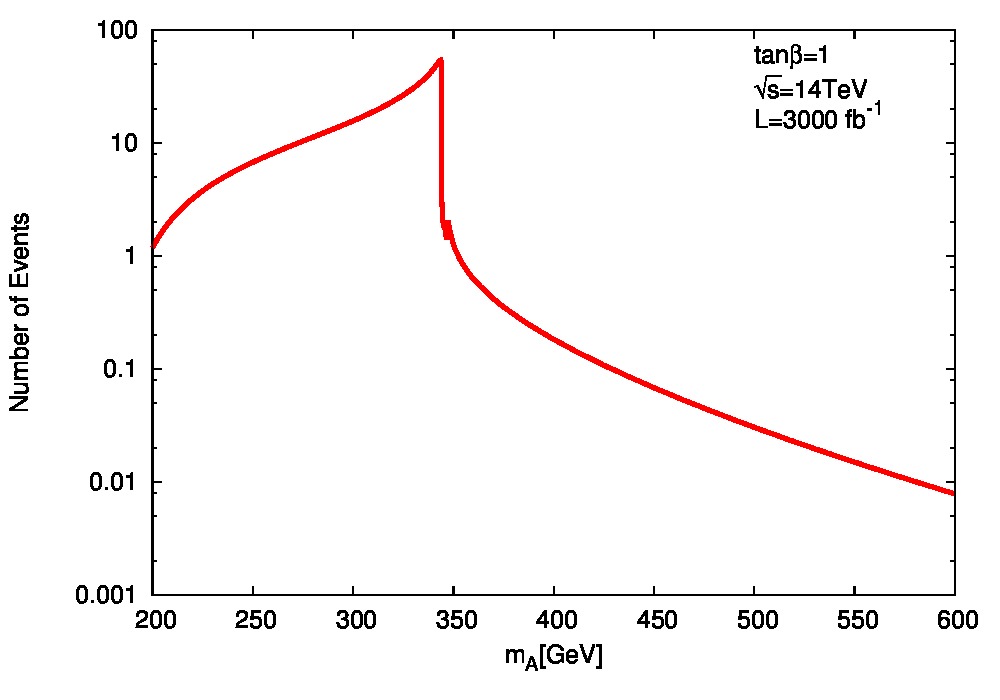}\includegraphics[width=0.45\textwidth]{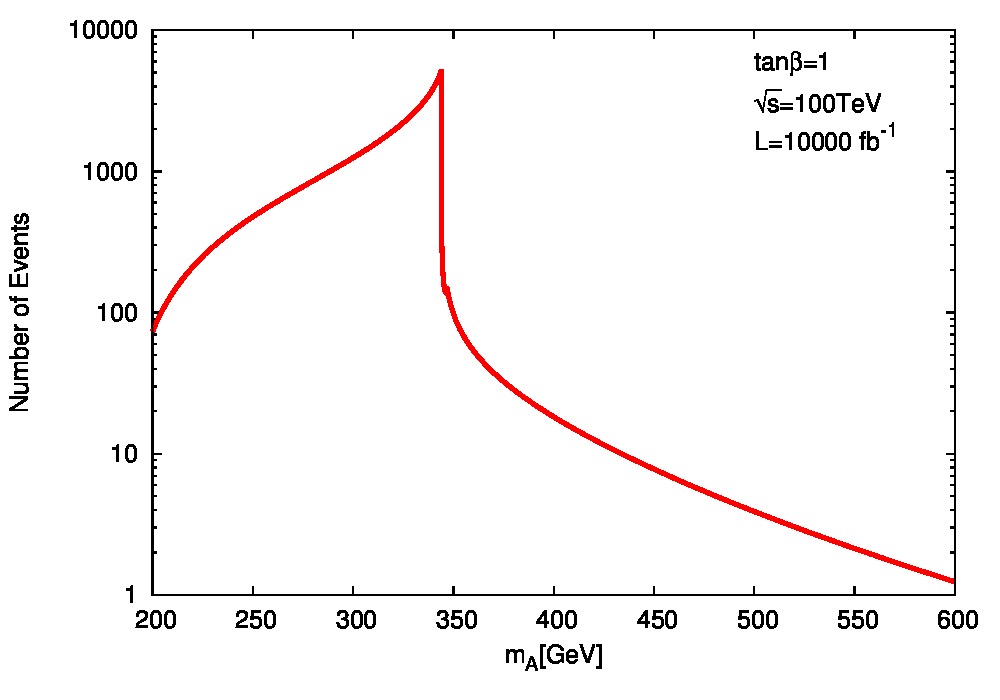}
\caption{Number of events obtained at the 14 TeV LHC with a total integrated luminosity of 3000 $fb^{-1}$ (left) and at a 100 TeV machine 
with 10 ab$^{-1}$ integrated luminosity for a 2HDM of any type (as $\tan \beta =1$). We assume that the efficiency factor is $\epsilon\approx 0.35$.}
\label{fig-new-plot}
\end{figure}

\begin{figure}[h!]\centering
\includegraphics[width=0.45\textwidth]{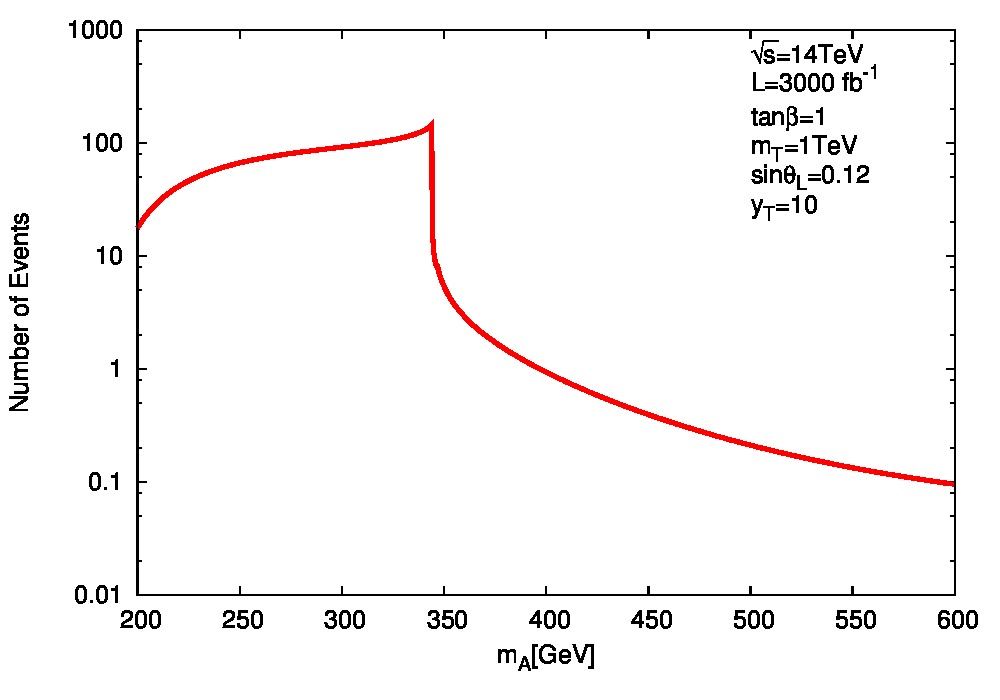}\includegraphics[width=0.45\textwidth]{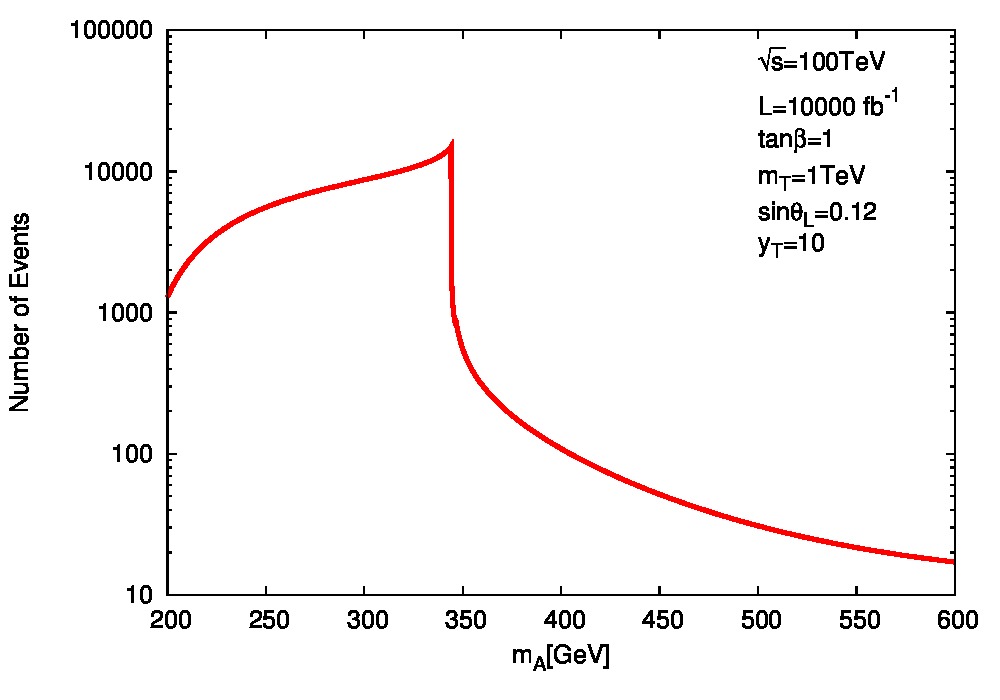}
\caption{Number of events obtained at the 14 TeV LHC with a total integrated luminosity of 3000 $fb^{-1}$ (left) and at a 100 TeV machine 
with 10 ab$^{-1}$ integrated luminosity for the 2HDM+T. We assume that the efficiency factor is $\epsilon\approx 0.35$.}
\label{fig-new-plot2}
\end{figure}

In Figure~\ref{fig-new-plot} we show the number of raw events eq.(\ref{rawevents}) for the 14 TeV LHC with 3000 $fb^{-1}$ integrated luminosity 
(left panel) and for a 100 TeV hadron machine with 10 $ab^{-1}$ luminosity (right panel). 
It is clear that even with the high luminosity option at the LHC, the golden mode could only be probed, if at all, in a very narrow region  
of the 2HDM parameter space where $m_A\approx 2 m_t$ which is due to $t\bar{t}$ threshold effect. Furthermore, this number of events
is obtained for $\tan \beta = 1$, and above that value it is hopeless to expect any significant number of events.
Of course, if the bounds would allow to lower $\tan\beta$ to values below 1, the number of events would grow. However, present and future constraints
on the 2HDM are moving the parameter space further and further away from the small $\tan \beta$ values.
Moving to a 100 TeV machine it is clear that the golden mode could be probed with an integrated luminosity of 10 ab$^{-1}$ 
in the CP-odd mass range of 200 to 400 GeV. In this mass range the total number of events produced would be in the range
of roughly 100 to 6000 events. In the 2HDM+T (Figure~\ref{fig-new-plot2} ) the number of events is larger than for the 2HDM but again it is in the range of 200 to 400 GeV
that the model can be probed. Two final comments are in order. First, it is clear that at the end of the present LHC run and even more
at the end of the high luminosity phase, the 2HDM+T will be very similar to the 2HDM, because if nothing is found, the bound on the 
vector top will grow. At the same time also the 2HDM will be closer to the alignment limit. Second, if a pseudoscalar  is in the mass
range of 200 to 400 GeV it will certainly be first discovered in another channel.

Here we should add that although there is no estimate, the 2HDM-T will be closer and closer to the 2HDM if nothing is found
because the mass of the new vector quark will have to be larger and larger.

\section{Mimicking a CP-violating 2HDM} 

In this work we are considering a scenario where $\sin (\beta -\alpha) =1$ (the alignment limit),
the lightest scalar is the SM-like Higgs with a mass $m_h = 125$ GeV and $m_{H^+}=m_{H}=600$ GeV.
In this scenario the pseudoscalar will decay mainly to fermions. In the right panel of figure~\ref{2hdmt100tev}
we present the partial width $\Gamma (A \to ZZ)$ as a function of $m_A$ for the Type I model for
values of $\tan \beta$ between 1 and 40.
We see that below the $t \bar t$ threshold, where $\sigma (pp \to A)\, BR(A \to ZZ)$ is largest,
the width is always below $10^{-5}$ GeV. In the alignment limit $\Gamma (H \to ZZ)$ is zero
at tree-level. However, using the prescription in~\cite{Krause:2016oke} it can be shown
that when the tree-level coupling $g_{HZZ}$ is zero the one loop $\Gamma (H \to ZZ)$ is 
of the order $10^{-5}$ to $10^{-4}$. This means that the $BR(A \to ZZ)$ and the $BR(H \to ZZ)$ 
will be of the same order of magnitude. Even if $BR(H \to ZZ)$ can be slightly larger, also
the production cross section of a pseudoscalar is larger than that of a scalar in gluon fusion.
Hence, if it is true that a 100 TeV collider will be able to probe very small
branching ratios of scalars to $ZZ$ it will most certainly be unable to tell the CP number
of the new scalar particles. Furthermore, in the alignment limit the branching ratio for 
the decay $A \to h Z$ is exactly zero. However, when moving away from this limit $BR (A \to hZ)$
will again be small but non-zero. So, if a scalar particle is detected can we probe its CP?

As discussed in the introduction, both the CP-conserving and the CP-violating
2HDM have the same Lorentz structure in the Higgs couplings to massive gauge bosons. Therefore,
angular distributions in the final states of $ZZ$ or $WW$ (four leptons) or in Higgs associated
production, will not distinguish CP-conserving from CP-violating extensions such as the 2HDM and C2HDM.
If more than one scalar is found then it is possible to search for signals of CP-violation at the
LHC with a combination of three decays. In reference~\cite{Fontes:2015xva} it was shown that 
three simultaneous decays of Higgs such as $h_2 \to h_1 Z$, $h_1 \to ZZ$ and $h_2 \to ZZ$
are a sign of CP-violation in any model. Also $h_1 \to ZZ$, $h_2 \to ZZ$ and  $h_3 \to ZZ$, 
where $h_1, h_2, h_3$ are generic Higgs bosons, is a clear signal of CP-violation in the 2HDM
except if there is a CP-odd state that decays to $ZZ$ with a significant rate. That is, the combination
of these three last decays can distinguish a CP-conserving 2HDM from a CP-violating one. 
Also the simultaneous processes  $pp \to Z h_1, \, Z h_ 2, \, Z h_3$ was shown to be a sign
of CP-violation in~\cite{Akeroyd:2001kt, Arhrib:2001pg}. 
Note however
that there are many models with 3 CP-even scalars that can decay to $ZZ$, like the singlet extension
or the 2HDM extended with a singlet. However, when the rate of $h_i \to ZZ$ becomes too small
we no longer know if this is just a very suppressed tree-level process or one that only appears at
one-loop, as happens for $A \to ZZ$. As discussed in the introduction, a different manifestation
of the same phenomena is through the measurement of anomalous $ZZZ$ couplings. At the moment
there are no predictions for these measurements at 100 TeV. However, as discussed in detail in the introduction,
if no new scalars are found, and because the measurement of $f_4$ is still orders of magnitude away from
the maximal calculated values in extensions of the SM, it should be clear that processes
involving Higgs and gauge bosons either at tree-level or in loops cannot be used to probe the CP of the scalars. 

In conclusion, if a new scalar (and only one) is found at these very low
rates, great precision is needed both from the experimental side and from the theoretical side
with the calculation of higher order corrections. But more important is that it becomes imperative 
to study the Higgs Yukawa couplings of the models. While in the 2HDM the Yukawa couplings
are either just a constant (scalar) or proportional to $\gamma_5$ (pseudoscalar), in the C2HDM
the Yukawa couplings have the form  $g^{h \bar f_i f_i}_{SM} \, (a_i + i b_i \gamma_5)$, where $g^{h  \bar f_i f_i}_{SM}$ is the SM coupling
for fermion $i$, and $i=U,D,L$, ($U,\, D, \, L $ stand for up-quarks, down-quarks and lepton couplings). Let us  define the angle that measures the relative strength of pseudoscalar to scalar Yukawa component, $\phi_i$, as
\begin{equation}
\tan \phi_i  = b_i/a_i \qquad i=U,\, D, \, L \, ,
\end{equation}
which could in principle be measured in direct experiments at the LHC. 
The ATLAS and CMS collaborations have not started any direct measurements of CP-violation despite many proposals from 
the theoretical community. These proposals all have in common the need for high or even very high luminosity at the LHC
and focus mainly on the  $tth$ and on the $\tau^+ \tau^- h$ couplings.
 Measurement of $b_U/a_U$ were proposed in the process $pp \to t \bar t h$
in~\cite{Gunion:1996xu} and several proposals followed as for instance the ones in~\cite{Boudjema:2015nda, Santos:2015dja}.
These use CP-odd variables together with asymmetries. Other proposals to probe the CP-nature of a scalar
in the $tth$ vertex include the process $pp \to hjj$~\cite{DelDuca:2001fn} as first proposed
in~\cite{Field:2002gt} and again more recently in~\cite{Dolan:2014upa},
where an exclusion of $\phi_t > 40^o $ ($\phi_t > 25^o$) for an integrated luminosity of 50 fb$^{-1}$ (300 fb$^{-1}$) was obtained for 14 TeV
and assuming $\phi_t=0$ as the null hypothesis. There are also a number of studies for the $\tau^+ \tau^- h$
vertex~\cite{Berge:2008wi, Harnik:2013aja, Askew:2015mda}
and a detailed study taking into account the main backgrounds~\cite{Berge:2014sra, Berge:2015nua} lead to an estimate in the precision
of $\Delta \phi_\tau$ of $15^o$ ($9^o $) for a luminosity of 150 fb$^{-1}$
(500 fb$^{-1}$) and a center of mass energy of 14 TeV. These studies show that even for a Higgs with a mass of 125 GeV
and SM-like couplings very high luminosities are needed.
An educated guess~\footnote{Werner Bernreuther, private communication.} for a 100 TeV pp collider would be to attain
$\Delta \phi_\tau \sim 1^o$ to $2^o$  with
an integrated luminosity of 10 ab$^{-1}$. The crucial thing is to keep control of the systematic
uncertainties.
Still, it is expectable that for a heavier Higgs the prospects will be 
much worse even if only because of the lower production cross section.


\section{Conclusions}

In this work we have analysed the detection of a pseudoscalar produced 
in gluon fusion plus $b \bar b$ initiated process and decaying to a pair of gauge bosons. We worked in the
alignment limit of the 2HDM, where $\sin (\beta - \alpha) = 1$, driven by the precision measurements of the Higgs 
couplings that have shown that one of the scalars has to resemble the SM Higgs.  
The experimental search for a generic scalar particle at the LHC Run 1
has been already performed by ATLAS and CMS. A small portion of parameter 
space of the 2HDM has already been probed in the 
search with two-photons in the final state. For the case of the final states
with two massive gauge bosons we are still at least one order of magnitude 
away from highest possible rates in the model.

We have also analysed the 2HDM model with an extra vector like quark, 2HDM+T.
Due to the extra loop contribution from the top partner we can have an 
enhancement of both the production cross section and of the decay widths. In fact
we have shown that the results for $pp \to A \to \gamma \gamma$ already
exclude a substancial region of the parameter space below the $t \bar t$ threshold.
As for the decays to massive vector bosons the enhancement is not enough
to reach the exclusion limit obtained during Run 1.
   
We have shown that in a future 100 TeV collider with a luminosity
of 10-20 ab$^{-1}$ almost all parameter space of four 2HDM Yukawa types will be probed
in the case of the decay $A \to \gamma \gamma$. However, for the pseudoscalar decays
into massive gauge bosons, possibly only a small portion of the parameter space will be at experimental reach.
Also it is important to note that when all rates in the $ZZ$ final state become
very small it will be extremely hard to use them to search for CP-violation.
In the case of the 2HDM+T as the rates are much higher it is expected that a larger
region of the parameter space will be probed. By the end of Run 2, and if no new physics is found, the model will be closer
to alignment, and the limits on heavy scalars will be stronger. In that case, the 100 TeV collider will start operation
with severe constraints both on the Higgs couplings and on the masses/couplings of extra scalars. Note that in the limit where the new top
decouples, the results are similar to the 2HDM ones.

One of the main ideas that triggered this work was the search for CP-violation. In the CP-violating 2HDM, the decays of any scalar to $ZZ$
are allowed. However, a CP-odd particle cannot decay to $ZZ$ at tree-level. We have shown that for the 2HDM the process $A \to ZZ$ is of the same order
of magnitude as $H \to ZZ$ if the tree-level $HZZ$ coupling is zero, that is, if we are close to aligment. If such a final state is detected at Run 2 or at a future 100 TeV collider
with a very low rate, it will be very hard to conclude anything about CP-violation. If a new scalar is detected in this final state with a higher rate
it can then be a scalar from a 2HDM but also a pseudoscalar from an extended version of the 2HDM, the 2HDM+T. Hence, and fortunately,
a lot of work is expected to pinpoint the underlying model. The CP-nature of any new scalar will for sure have to rely on direct measurements
of the ratio of pseudoscalar to scalar components in the Yukawa couplings. Very preliminary studies have been performed for a scalar
decaying into $\tau^+ \tau^-$ and in $tth$ production.

Finally we note that many other models with an extended Higgs sector will behave exactly like the 2HDM. In fact,
if we extend the 2HDM with a singlet we end up with a model that in the alignment limit has a pseudoscalar that
couples to the remaining SM particles exactly like the 2HDM. Therefore, our conclusions are valid for all extensions
of the SM where alignment leads to a pseudoscalar with 2HDM-like couplings.

\vspace*{0.05cm}
\section*{Acknowledgments}
The authors are supported by the grant H2020-MSCA-RISE-2014 no. 645722
(NonMinimalHiggs).
This work is also supported by the Moroccan Ministry of Higher
Education and Scientific Research MESRSFC and  CNRST: Projet PPR/2015/6. J.E would like to thank Shaaban Khalil 
for the hospitality extended to him during his stay in the Center for Fundamental Physics (CFP) at 
Zewail City of Science and Technology where part of this work has been done. He also 
acknowledges the receipt of the grant from the Abdus Salam International Center for Theoretical Physics, Trieste, Italy.
R.S. is also supported in part by the National Science Centre, Poland, the HARMONIA project under contract UMO-2015/18/M/ST2/00518.
We acknowledge discussions with Nikos Rompotis, Werner Bernreuther and Ritesh Singh.

\vspace*{0.05cm}
\bibliographystyle{h-physrev}
\bibliography{AtoVVfinal}
\end{document}